\documentclass[manuscript]{acmart}
\usepackage[colorinlistoftodos]{todonotes}
\usepackage{amsmath,amsfonts}

\usepackage[most]{tcolorbox}
\usepackage{amssymb}
\usepackage{mathtools}
\usepackage{algorithmic}
\usepackage{algorithm}
\usepackage{float,subfloat}
\usepackage{flafter} 
\usepackage{graphicx}
\usepackage{textcomp}
\usepackage{xcolor}
\usepackage{colortbl}
\definecolor{LightGray}{gray}{0.9}
\usepackage{listings}
\usepackage{amsthm}
\usepackage{enumitem}
\usepackage{hyperref}
\usepackage{booktabs}
\usepackage{siunitx}
\usepackage{threeparttable,multicol,makecell,multirow,bm,adjustbox,booktabs,mdframed}
\usepackage{placeins}
\usepackage{tcolorbox}
\tcbuselibrary{listings}
\usepackage[normalem]{ulem} 
\tcbuselibrary{listingsutf8}
\newtcolorbox{rqbox}[1]{
  colback=gray!5,
  colframe=black!60,
  boxrule=0.5pt,
  arc=2pt,
  left=6pt,
  right=6pt,
  top=6pt,
  bottom=6pt,
  title=\textbf{#1},
  fonttitle=\bfseries
}
\newcommand{\tech}{\textsc{CAT}} 
\newcommand{\techwithpath}{\textsc{CAT}\textsubscript{path}} 
\newcommand{\techchain}{\textsc{CAT}\textsubscript{chain}} 
\newcommand{\techdep}{\textsc{CAT}\textsubscript{dep}} 
\newcommand{\techvariant}{\textsc{CAT}\textsubscript{vanilla}}
\newcommand{\panta}{\textsc{PANTA}}
\newcommand{\pantavariant}{\textsc{PANTA}\textsubscript{tb}} 

\newcommand{\est}[1]{#1} 

\newif\ifshowchanges
\showchangesfalse

\definecolor{RevBlue}{HTML}{1A5FB4}

\ifshowchanges
  \newcommand{\added}[1]{\textcolor{RevBlue}{#1}}        
  \newcommand{\deleted}[1]{\textcolor{red}{\sout{#1}}}   
  \newenvironment{addedblock}{\color{RevBlue}}{}         
\else
  \newcommand{\added}[1]{#1}
  \newcommand{\deleted}[1]{}
  
  \newenvironment{addedblock}{}{}
\fi


\AtBeginDocument{%
  }

\setcopyright{acmlicensed}
\copyrightyear{2018}
\acmYear{2018}
\acmDOI{XXXXXXX.XXXXXXX}
\acmConference[Conference acronym 'XX]{Make sure to enter the correct
  conference title from your rights confirmation email}{June 03--05,
  2018}{Woodstock, NY}
\acmISBN{978-1-4503-XXXX-X/2018/06}

\begin{document}

\title{Call-Chain-Aware LLM-Based Test Generation for Java Projects}

\author{Guancheng Wang}
\email{guancheng.wang@ul.ie}
\orcid{0000-0002-4338-8813}
\affiliation{%
  \institution{Research Ireland Lero Centre for Software and University of Limerick}
  \city{Limerick}
  \state{Limerick}
  \country{Ireland}
}

\author{Qinghua Xu}
\email{qinghua.xu@ul.ie}
\orcid{0000-0001-8104-1645}
\affiliation{%
  \institution{Research Ireland Lero Centre for Software and University of Limerick}
  \city{Limerick}
  \country{Ireland}
}

\author{Lionel C. Briand}
\email{lionel.briand@ul.ie}
\orcid{0000-0002-1393-1010}
\affiliation{%
  \institution{Research Ireland Lero Centre for Software and University of Limerick}
  \city{Limerick}
  \country{Ireland}
}
\affiliation{
    \institution{University of Ottawa}
    \city{Ottawa}
    \country{Canada}
}

\author{Zhaoqiang Guo}
\email{gzq@smail.nju.edu.cn}
\orcid{0000-0001-8971-5755}
\affiliation{
    \institution{The State Key Laboratory of Blockchain and Data Security, Zhejiang University}
    \city{Hangzhou}
    \country{China}
}

\author{Kui Liu}
\email{brucekuiliu@gmail.com}
\orcid{0000-0003-0145-615X}
\affiliation{
    \institution{Software Engineering Application Technology Lab, Huawei}
    \city{Hangzhou}
    \country{China}
}
\renewcommand{\shortauthors}{Wang et al.}

\begin{abstract} 
Large language models (LLMs) have recently shown strong potential for generating project-level unit tests. However, existing state-of-the-art approaches primarily rely on execution-path information to guide prompt construction, which is often insufficient for complex software systems with rich inter-class dependencies, deep call chains, and intricate object initialization requirements. In this paper, we present CAT, a novel call-chain-aware LLM-based test generation approach that explicitly incorporates call-chain and dependency contexts into prompts through dedicated static analysis. To construct executable, semantically valid test contexts, CAT systematically models caller--callee relationships, object constructors, and third-party dependencies, and supports iterative test fixing when generation failures occur. We evaluate CAT on the widely used Defects4J benchmark and on four real-world GitHub projects released after the LLM's cut-off date, and compare it with the state-of-the-art LLM-based approach PANTA and the search-based tool EvoSuite. \added{The results show that, across projects in Defects4J, CAT improves line coverage, branch coverage, and mutation score by 12.52, 12.97, and 14.39 percentage points (18.02\%, 21.52\%, and 26.94\% relative), respectively, over PANTA, and by 9.67, 4.17, and 15.70 percentage points (13.37\%, 6.04\%, and 30.13\% relative) over EvoSuite. On the post-cutoff projects, CAT also achieves the highest average line coverage, branch coverage, and mutation score among all compared approaches.} An ablation study further demonstrates the importance of call-chain and dependency contexts in CAT. 
\end{abstract}



\received{20 February 2007}
\received[revised]{12 March 2009}
\received[accepted]{5 June 2009}

\maketitle

\section{Introduction} \label{sec:intro}
Unit tests play a fundamental role in ensuring software quality and reliability. However, writing high-quality unit tests remains labor-intensive and requires substantial domain knowledge, posing a long-standing challenge for software engineers~\citep{fraser2011evosuite,tillmann2008pex,tufano2019empirical,pacheco2007randoop}. To mitigate this burden, a variety of automated test generation techniques have been proposed, including random testing (e.g., Randoop~\citep{pacheco2007randoop}), search-based testing (e.g., EvoSuite~\citep{fraser2011evosuite}), and symbolic execution–based approaches (e.g., KLEE~\citep{cadar2008klee} and Pex~\citep{tillmann2008pex}). These techniques have proven effective at detecting critical defects by systematically exploring program behaviour.
\added{These techniques differ in how they obtain test inputs. Random testing extends random sequences of method calls using execution feedback; search-based testing evolves test suites towards coverage goals; and symbolic execution derives inputs by solving path constraints. Their test oracles are typically inferred from observed executions or from generic properties (e.g., crashes and contracts). KLEE and Pex target C and .NET programs, respectively, whereas Randoop and EvoSuite focus on unit test generation for Java---the setting studied in this paper. We include EvoSuite as a non-LLM baseline in our evaluation (Section~\ref{sec:eval}) because it is commonly used as the reference non-LLM tool in recent studies of LLM-based test generation~\citep{tang2024chatgpt,yang2024evaluation,wang2024hits,pan2025aster,abdullin2025test} and has outperformed Randoop in coverage, mutation score, and real-fault detection~\citep{devroey2023juge,shamshiri2015automatically,almasi2017industrial}.}
More recently, advances in machine learning and large language models (LLMs) have further pushed the frontier of automated test generation~\citep{wang2024hits,pan2025aster,nan2025test}. LLM-based approaches leverage learned programming knowledge and contextual reasoning to generate readable, functional test cases, significantly improving both the automation and effectiveness of testing.
Despite these advances, most existing work focuses on method-level~\citep{wang2025mutation,xu2025hallucination,ryan2024code} or class-level~\citep{gu2025llm,zhang2024testbench} test generation, which typically considers only the program logic within a single module. In contrast, project-level test generation, which must account for complex interprocedural dependencies, execution contexts, and interactions across multiple modules, remains largely underexplored and challenging. This gap limits the applicability of current techniques in large-scale software systems.

State-of-the-art project-level test generation approaches leverage execution-path feedback to iteratively guide LLMs toward uncovered code regions, including \panta{}~\citep{gu2025llm}, PALM~\citep{wu2025generating}, and SymPrompt~\citep{ryan2024code}. These approaches dynamically collect uncovered execution paths or path constraints to guide test generation toward higher coverage. While such strategies have demonstrated promising results in certain settings, they rely on incorporating uncovered execution paths into the prompt as feedback signals and generally do not explicitly model dependencies across modules and execution contexts, as illustrated in Listing~1, where inter-class dependencies are required to correctly invoke the focal method but are not captured by existing methods. As a result, they often fail to generate valid tests for projects with complex inter-module dependencies.
Empirical results in Section~\ref{sec:eval} further confirm this limitation. Using \panta{} as a state-of-the-art representative of this line of work, we observe that on the JxPath project from the Defects4J~\citep{just2014defects4j} dataset--where methods heavily depend on external dependencies (e.g., dependencies on other modules, third-party libraries)--it achieves only 56.97\% line coverage and 53.88\% branch coverage.

This observation suggests that the key difficulty lies not in identifying uncovered code locations but in constructing valid execution contexts, such as cross-class dependency setups, object initializations, and call sequences required to invoke the focal method. However, existing path-guided approaches provide limited insight into how such contexts are formed, as they focus primarily on uncovered code feedback without modeling dependency structures or object initialization processes.
We further observe that the required execution contexts are inherently encoded in the call chains leading to the focal method, together with the constructors of the involved classes. These call chains capture how objects are created, how external dependencies are invoked, and the order in which different components interact.
Therefore, we propose \tech{}, built on a dedicated static analysis framework that systematically captures relevant execution contexts. Specifically, \tech{} extracts call-chain information and associated initialization contexts, including caller-callee relationships, object constructions, and third-party dependencies, and explicitly provides the LLM with such rich project-level context, enabling it to generate valid tests even for methods with complex inter-module and third-party dependencies.

Using the same benchmark (Defects4J) as in its original experiments, \panta{} is, on average, outperformed by \tech{} by 12.52, 12.97, and 14.39 percentage points (18.02\%, 21.52\%, and 26.94\%) in line coverage, branch coverage, and mutation score, respectively. As discussed later in Section~\ref{sec:eval}, \tech{} generates tests for each focal method individually, whereas \panta{} generates tests once per focal class. As a result, \tech{} may incur higher overall generation costs due to its finer-grained generation process. Thus, we further perform a fairer comparison giving \panta{} the same time budget as \tech{}, where the latter still achieves improvements of 8.94, 9.61, and 14.76 percentage points (12.23\%, 15.11\%, and 27.83\% relative) in line coverage, branch coverage, and mutation score over the resulting time-bounded variant of \panta{}. \added{We also compare \tech{} against EvoSuite, a widely used search-based test generation tool that does not rely on an LLM. \tech{} outperforms EvoSuite by 9.67, 4.17, and 15.70 percentage points (13.37\%, 6.04\%, and 30.13\% relative) in line coverage, branch coverage, and mutation score, respectively; however, EvoSuite attains higher branch coverage on five of the 14 projects.}

To further assess the generalizability of our approach and increase the realism of our comparisons, we evaluate \tech{} on four additional Java projects collected from GitHub that were released after the employed LLM became publicly available, thereby making them unlikely to have been part of its training data. As discussed in Section~\ref{sec:subjects}, these projects are also more complex than those in Defects4J. On these unseen projects, \tech{} improves line coverage, branch coverage, and mutation score by 13.58, 12.02, and 24.64 percentage points (25.09\%, 25.91\%, and 94.37\% relative), respectively, over the original \panta{}, and by 7.80, 7.48, and 19.88 percentage points (13.02\%, 14.68\%, and 64.40\%) over its time-bounded variant. \added{Compared with EvoSuite, \tech{} improves the three metrics by 23.19, 14.16, and 17.30 percentage points (52.10\%, 31.99\%, and 51.72\%), respectively.}
These results suggest that explicitly modeling execution context, including call chains, constructor information, and third-party dependencies, substantially improves both the coverage and the fault-detection capability of LLM-generated tests in project-level test generation, also on previously unseen projects.

In summary, this paper makes the following contributions:
\begin{itemize}
\item We identify a key limitation of existing path-guided project-level test generation approaches: uncovered execution paths alone are insufficient for constructing valid execution contexts in projects with complex inter-module and third-party dependencies. To address this, we introduce a context modeling strategy and provide a dedicated static analysis framework to systematically capture relevant execution contexts, including call chains, object constructors, and third-party dependencies.
\item Based on this strategy, we propose \tech{}, an LLM-based project-level test generation approach. To the best of our knowledge, \tech{} is the first approach to explicitly incorporate call-chain- and dependency-aware execution contexts into LLM-based project-level test generation, enabling effective test generation for complex software systems.
\item We conduct an extensive empirical evaluation on Defects4J and four newly released GitHub projects that are unseen by the employed LLM and exhibit higher complexity than Defects4J in terms of inter-class dependencies and project structure, demonstrating substantial improvements in line coverage, branch coverage, and mutation score over \panta{}, the state-of-the-art LLM-based baseline, and over the search-based tool EvoSuite, and we release an open-source toolkit of \tech{}.
\end{itemize}

The remainder of this paper is organized as follows. Section~\ref{sec:motivate} presents the motivating example for this paper. Section~\ref{sec:approach} presents the design of our approach \tech{}. Section~\ref{sec:eval} describes the experimental setup and reports the evaluation results. Section~\ref{sec:related} reviews related work. Finally, Section~\ref{sec:conclusion} concludes the paper.

\section{Motivating Example} \label{sec:motivate}
\definecolor{keywordcolor}{RGB}{0,0,180}
\definecolor{commentcolor}{RGB}{0,128,0}
\definecolor{stringcolor}{RGB}{163,21,21}

\lstdefinestyle{paperlisting}{
    language=Java,
    basicstyle=\ttfamily\footnotesize,
    keywordstyle=\color{keywordcolor}\bfseries,
    commentstyle=\color{commentcolor}\itshape,
    stringstyle=\color{stringcolor},
    numbers=left,
  numberstyle=\tiny\color{black!60},
    stepnumber=1,
  numbersep=8pt,
    showspaces=false,
    showstringspaces=false,
    showtabs=false,
    frame=single,
  framerule=0.4pt,
  rulecolor=\color{black!35},
  backgroundcolor=\color{LightGray!20},
  framesep=5pt,
  xleftmargin=1.2em,
  xrightmargin=0.4em,
    breaklines=true,
  breakatwhitespace=false,
  tabsize=4,
  columns=fullflexible,
  keepspaces=true,
  captionpos=b
}

\lstdefinestyle{compactlisting}{
  style=paperlisting,
  basicstyle=\ttfamily\scriptsize,
  numbersep=6pt,
  xleftmargin=1em,
  xrightmargin=0.2em
}

\lstset{
  style=paperlisting
}

In this section, we present a motivating example to illustrate the limitations of existing approaches in
handling focal classes with \textit{external dependencies}.

The first two focal methods shown in Listing~\ref{lst:xml-factory}, \lstinline|hasXMLFormat| and \lstinline|hasFormat|, from the \lstinline|XmlFactory| class in the JacksonXml-5f project of the Defects4J dataset, illustrate a common scenario in real-world codebases. Both their input type \lstinline|InputAccessor| and return type \lstinline|MatchStrength| are defined outside \lstinline|XmlFactory|, introducing dependencies on external classes and requiring non-trivial object construction and cross-class interaction to correctly invoke these methods. 
Such scenarios highlight a fundamental challenge for coverage-driven test generation approaches. Since these methods depend on externally defined types and initialization logic, simply guiding test generation toward uncovered code regions is insufficient. In particular, when the generation process is primarily driven by coverage feedback, it tends to focus on exploring new execution paths without explicitly reasoning about how required objects are constructed or how inter-class dependencies are resolved. As a result, the generated tests often fail to establish valid execution contexts for methods with complex dependencies, limiting their ability to reach and execute the target methods effectively.

In addition, focal classes often contain \textit{overloaded methods}, such as multiple \lstinline|configure| methods with different parameters (the last two methods as shown in Listing~\ref{lst:xml-factory}). 
Method overloading is a common feature of real-world Java codebases, further complicating method-level test generation, as different overloads may require distinct input constraints and execution contexts.
Empirical evidence~\citep{khan2024empirical,gil2010use} shows that method overloading is non-trivial in real-world Java code:
\begin{itemize}
    \item In a dataset of open-source Java repositories, more than 5\% of method declarations are overloaded on average, with some repositories reaching 27\%~\citep{khan2024empirical}.
    \item An earlier study~\citep{gil2010use} found that over 14\% of methods in a large Java corpus are overloaded.
\end{itemize}

Our approach addresses both limitations through a dedicated static analysis framework. First, we generate tests one focal method at a time, which avoids ambiguity caused by overloaded methods. Second, we systematically extract relevant execution contexts, including both call-chain information and external class dependencies, via static analysis (see Listings~\ref{lst:execution-path} and \ref{lst:dependency-context}).

We illustrate our approach using the focal method \lstinline|hasFormat| as an example. 
\added{Listing~\ref{lst:execution-path} shows the call-chain context of \lstinline|hasFormat| in a simplified form. Each block gives the declaring class, the method with its parameter types and, after the arrow, its return type. For brevity, we list each indirect call chain from its public entry method up to the last method before the focal method; the final call to the focal method (which ends every chain) is omitted.}

\added{The first block is such an indirect chain. \lstinline|DataFormatDetector.findFormat(byte[])| wraps the given bytes in an \lstinline|InputAccessor| and passes it to the private helper \lstinline|_findFormat|, which calls \lstinline|hasFormat| on the \lstinline|JsonFactory| instances registered in the detector until one of them reports a sufficiently strong match. Since \lstinline|XmlFactory| extends \lstinline|JsonFactory| and overrides \lstinline|hasFormat|, CHA links this call to the focal method, which is reached at run time when the detector holds an \lstinline|XmlFactory|. The private helper cannot be called from a test and is removed by the visibility filter (Section~\ref{sec:path-filtering}); thus, \lstinline|findFormat| is both the entry method and the last method before the focal method. The second block is the focal method itself, which, being public, can also be invoked directly.}

Listing~\ref{lst:dependency-context} shows the extracted dependency context. For each related class, we collect all public constructors. For abstract classes (e.g., \lstinline|InputAccessor|), we further identify their concrete implementations. This information enables the LLM to construct valid object instances required for executing the focal method and its callers.

To improve coverage of the focal method, we prompt the LLM to generate tests that either directly invoke the focal method or indirectly invoke it through call-chain methods. This guidance helps the LLM construct diverse inputs and execution paths, thereby triggering more branches within the focal method.

For example, to indirectly invoke \lstinline|hasFormat| via \lstinline|findFormat|, additional dependencies—such as instantiating a \lstinline|DataFormatDetector| object—must be satisfied. Leveraging the extracted dependency context shown in Listing~\ref{lst:dependency-context}, the LLM successfully generates both indirect tests (Listing~\ref{lst:indtest}) and direct tests (Listing~\ref{lst:dtest}) for the focal method.

By incorporating both dependency context and call chain information, our approach generates tests that effectively cover the focal methods in Listing~\ref{lst:xml-factory}, resulting in 23.49\% and 46.16\% increases in line and branch coverage, respectively, compared to \panta{}, which serves as a representative of the most recent state-of-the-art path-guided test generation approaches.

\begin{lstlisting}[
  caption={Example focal methods in XmlFactory exhibiting dependencies on external classes},
  label={lst:xml-factory},
  style=paperlisting,
  escapeinside={(*@}{@*)}
]
com.fasterxml.jackson.dataformat.xml.XmlFactory:
    - public static MatchStrength hasXMLFormat((*@\textbf{\texttt{InputAccessor}}@*) acc)
    - public MatchStrength hasFormat((*@\textbf{\texttt{InputAccessor}}@*) acc)
    - public final XmlFactory configure(ToXmlGenerator.Feature f, boolean state)
    - public final XmlFactory configure(FromXmlParser.Feature f, boolean state)
\end{lstlisting}

\begin{lstlisting}[
  caption={Call-chain context of the focal method \texttt{XmlFactory.hasFormat}},
  label={lst:execution-path},
  style=paperlisting
]
Call-chain context:
  com.fasterxml.jackson.core.format.DataFormatDetector
    #findFormat(byte[])
      -> com.fasterxml.jackson.core.format.DataFormatMatcher

  com.fasterxml.jackson.dataformat.xml.XmlFactory
    #hasFormat(com.fasterxml.jackson.core.format.InputAccessor)
      -> com.fasterxml.jackson.core.format.MatchStrength
\end{lstlisting}

\begin{lstlisting}[
  caption={Dependency context of \texttt{XmlFactory.hasFormat}},
  label={lst:dependency-context},
  style=compactlisting
]
Dependency Context:

com.fasterxml.jackson.dataformat.xml.XmlFactory (public):
  - XmlFactory()
  - XmlFactory(javax.xml.stream.XMLInputFactory)
  - XmlFactory(javax.xml.stream.XMLInputFactory,
               javax.xml.stream.XMLOutputFactory)
  - XmlFactory(com.fasterxml.jackson.core.ObjectCodec)
  - XmlFactory(com.fasterxml.jackson.core.ObjectCodec,
               javax.xml.stream.XMLInputFactory,
               javax.xml.stream.XMLOutputFactory)

com.fasterxml.jackson.core.format.InputAccessor (public abstract):
  - no public constructors found
  - Known implementations:
      * InputAccessor$Std
      * DataFormatReaders$AccessorForReader

com.fasterxml.jackson.core.format.DataFormatDetector (public):
  - DataFormatDetector(com.fasterxml.jackson.core.JsonFactory[])
  - DataFormatDetector(java.util.Collection)

com.fasterxml.jackson.core.JsonFactory (public):
  - JsonFactory()
  - JsonFactory(com.fasterxml.jackson.core.ObjectCodec)
  - Known implementations:
      * XmlFactory
      * MappingJsonFactory
\end{lstlisting}


\begin{minipage}[t]{0.48\textwidth}
\begin{lstlisting}[style=compactlisting, numbers=left, caption={Test via DataFormatDetector}, label={lst:indtest}]
@Test
public void testHasFormatViaDataFormatDetector() throws IOException {
    String xmlContent = "<root><child>value</child></root>";
    byte[] bytes = xmlContent.getBytes();
    JsonFactory jsonFactory = new JsonFactory();
    DataFormatDetector detector = new DataFormatDetector(jsonFactory);
    DataFormatMatcher matcher = detector.findFormat(bytes);
    assertNotNull(matcher);
}
\end{lstlisting}
\end{minipage}%
\hfill
\begin{minipage}[t]{0.48\textwidth}
\begin{lstlisting}[style=compactlisting, numbers=right, caption={Test with mock InputAccessor},label={lst:dtest}]
@Test
public void testHasFormatWithValidXMLFixed() throws IOException {
    byte[] xmlBytes = "<root><child>text</child></root>".getBytes();
    InputAccessor mockAccessor = mock(InputAccessor.class);
    when(mockAccessor.hasMoreBytes()).thenReturn(true, true, false);
    when(mockAccessor.nextByte()).thenReturn((byte)'<').thenReturn((byte)'r').thenReturn((byte)'o');
    XmlFactory factory = new XmlFactory();
    MatchStrength result = factory.hasFormat(mockAccessor);
    assertEquals(MatchStrength.SOLID_MATCH, result);
}
\end{lstlisting}
\end{minipage}

\section{Approach} \label{sec:approach}
In this section, we introduce \tech{}, a \underline{C}all-chain-\underline{A}ware \underline{T}est-generation approach. Section~\ref{sec:overview} presents an overview of our approach, describing its iterative workflow consisting of a generation phase and a fixing phase, along with the prompts used in each phase. Section~\ref{sec:gen} presents the dedicated static analysis framework, detailing the algorithms for systematically extracting call-chain information and resolving dependency contexts to construct relevant execution contexts for test generation. Section~\ref{sec:genphase} describes the prompt design for the generation phase, which integrates the contexts extracted in Section~\ref{sec:gen}. Section~\ref{sec:fixphase} describes the prompt design for the fixing phase.

\subsection{Overview} \label{sec:overview}
\tech{} consists of two phases: a \textit{generation phase} and a \textit{fixing phase}. It can operate with or without existing tests. \added{When a test class already exists, \tech{} preserves the existing tests and inserts each newly generated test into the same class. A new test is retained only if the class still compiles and the new test passes. In this setting, the stopping criterion is initialized with the code coverage achieved by the existing tests.} When no tests are available, \tech{} first creates a pseudo test file with placeholder test functions, ensuring that newly generated tests can be compiled and executed within a valid test harness.

\begin{figure}[tbp]
    \centering
    \includegraphics[width=0.9\textwidth]{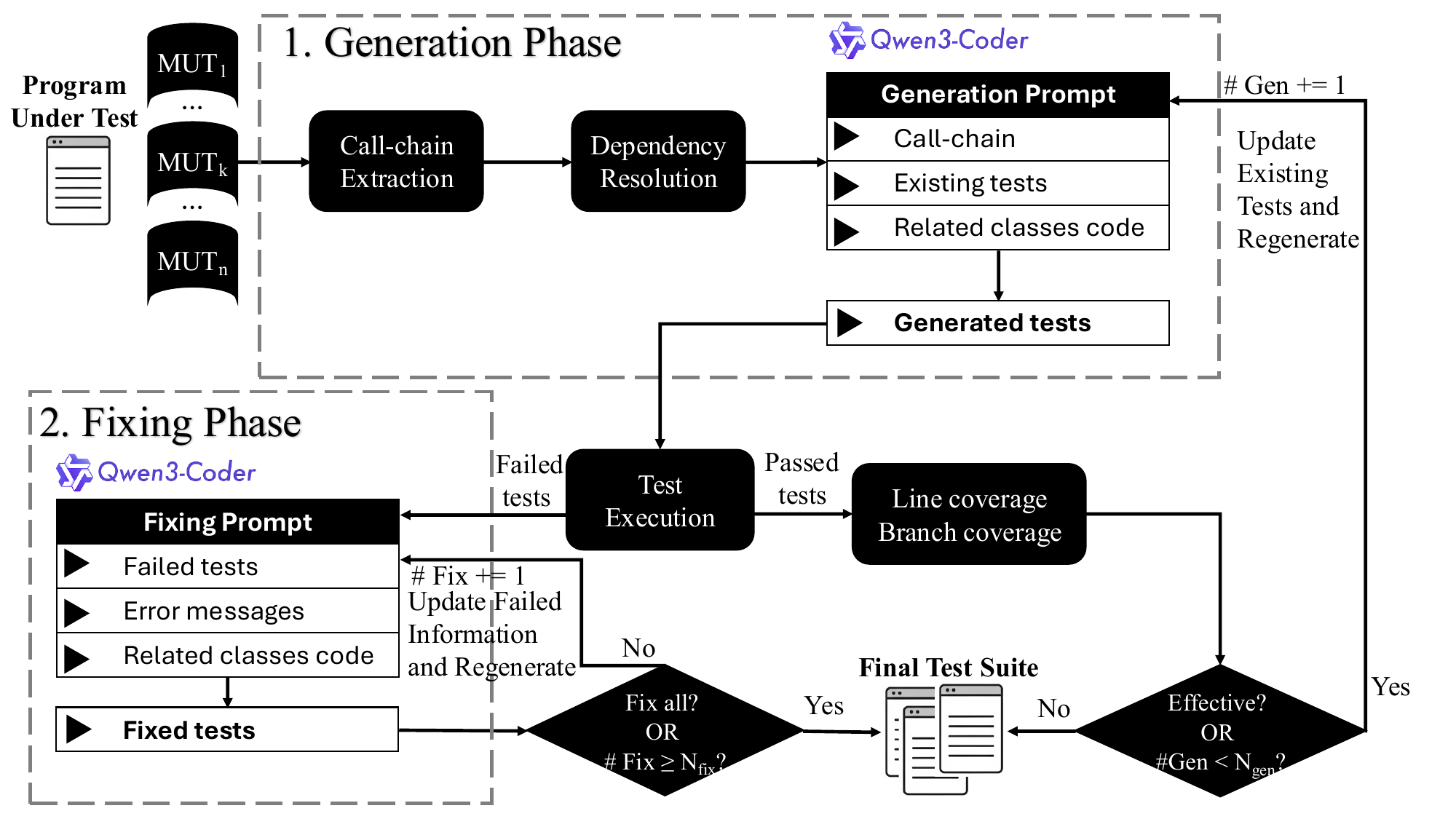}
    \caption{Overview of the \tech{} workflow}
    \label{fig:overview}
\end{figure}

As illustrated in Figure~\ref{fig:overview}, \tech{} follows an iterative workflow that alternates between these two phases. In the generation phase, \tech{} first statically extracts call-chain and dependency information for the method under test (MUT), which it performs only once before the iterative generation process. The extracted call-chain and dependency contexts, along with the source code of related classes, are incorporated into the generation prompt to guide test generation for candidate tests. Based on this prompt, \tech{} iteratively generates tests to maximize coverage. Section~\ref{sec:genphase} presents details of this phase.
In the fixing phase, \tech{} repairs failing tests by leveraging execution feedback. Specifically, failed tests and their corresponding error messages are incorporated into the fixing prompt, allowing up to \(N_{fix}\) repair attempts within each generation iteration. Section~\ref{sec:fixphase} presents details of this phase.
The workflow continues until the testing effort becomes ineffective, i.e., when neither line nor branch coverage improves for \(m\) consecutive iterations, or until the maximum number of generation iterations \(N_{gen}\) is reached, at which point the process terminates. 

\added{\tech{} generates regression tests: the current implementation serves as the reference for expected behavior. Assertions capture the values observed during execution, and the fixing phase updates failing assertions to match the observed behavior without modifying production code. A passing test therefore documents current behavior and flags future changes, but it does not establish that the behavior is correct. The same assumption underlies EvoSuite's regression assertions~\citep{fraser2011evosuite} and the tests retained by \panta{}~\citep{gu2025llm}.}

We do not present the full generation and fixing prompts in the paper because they are too long to include in the main text. Instead, detailed prompt templates are provided in the replication package. Nevertheless, as illustrated in the overview and further elaborated in the following sections, the generation prompt is carefully designed to incorporate extracted call-chain and dependency information. Next, we describe how we obtain these key components and integrate them into the prompt.

\begin{algorithm}[t]
\caption{Call Graph Construction and Metadata Collection}
\label{alg:cg-construction}
\begin{algorithmic}[1]

\REQUIRE Classpath $CP$, target class $C_t$, target method name $m_t$
\ENSURE Call graph $CG$, class visibility map $\textit{classVis}$, 
        method visibility map $\textit{methodVis}$, constructor map $\textit{ctors}$,
        implementation map $\textit{impls}$, entry points $\mathcal{R}$,
        target methods $\mathcal{T}$

\STATE \textbf{Phase 1: Analysis View Construction}
\STATE $V \gets \textsc{LoadBytecode}(CP)$
\STATE $\mathcal{C} \gets \textsc{GetClasses}(V)$

\STATE \textbf{Phase 2: Class and Method Metadata Collection}
\FOR{each class $c \in \mathcal{C}$}
    \STATE $\textit{classVis}[c] \gets \textsc{GetVisibility}(c)$
    \STATE $\textit{ctors}[c] \gets \textsc{GetConstructors}(c)$
    \FOR{each method $m \in c.\textit{methods}$}
        \STATE $\textit{methodVis}[m] \gets \textsc{GetVisibility}(m)$
    \ENDFOR
    \IF{$c$ is public}
        \STATE $\textit{impls}[c.\textit{superclass}].\textsc{Add}(c)$
        \FOR{each interface $i \in c.\textit{interfaces}$}
            \STATE $\textit{impls}[i].\textsc{Add}(c)$
        \ENDFOR
    \ENDIF
\ENDFOR

\STATE \textbf{Phase 3: CHA-Based Call Graph Construction}
\STATE $\mathcal{R} \gets \{m \mid m \in \mathcal{C}.\textit{methods} \land \textit{methodVis}[m] == \text{public}\}$
\STATE $CG \gets \textsc{ClassHierarchyAnalysis}(V, \mathcal{R})$
\STATE $\mathcal{T} \gets \{m \in C_t.\textit{methods} \mid m.\textit{name} == m_t\}$

\RETURN $CG$, $\textit{classVis}$, $\textit{methodVis}$, $\textit{ctors}$, $\textit{impls}$, $\mathcal{R}$, $\mathcal{T}$

\end{algorithmic}
\end{algorithm}

\subsection{Call Chain Extraction and Dependency Resolution} \label{sec:gen}

\tech{} is built on a dedicated static analysis framework that systematically captures relevant execution contexts for test generation. Specifically, it performs static analysis to identify potential invocation paths from accessible entry points to the target method and extracts call-chain information along these paths. It also analyzes class and method metadata to determine initialization requirements and related dependencies, including object constructors and third-party dependencies, thereby constructing executable, semantically valid contexts for test generation. 

As outlined in Algorithms~\ref{alg:cg-construction}, \ref{alg:path-extraction}, and \ref{alg:path-filtering}, this process consists of three main steps. First, we construct a call graph and collect relevant program metadata based on a Class Hierarchy Analysis (CHA~\citep{dean1995optimization}) (Algorithm~\ref{alg:cg-construction}). Next, we extract all invocation paths that can reach the target method under a bounded depth (Algorithm~\ref{alg:path-extraction}). 
Finally, we refine the extracted call paths to retain only feasible entry points for test generation, since some methods may not be publicly accessible or may belong to non-public or abstract classes that cannot be directly invoked in generated tests. Based on these filtered paths, we infer the set of classes that require initialization, along with their associated initialization methods, to ensure the generated tests can construct the necessary execution context (Algorithm~\ref{alg:path-filtering}).
We then incorporate the filtered call-chain paths and dependency information into the generation prompt to guide the test generation process, as described in Section~\ref{sec:overview}.

\subsubsection{Call Graph Construction and Metadata Collection}
\label{sec:cg-metadata}

As shown in Algorithm~\ref{alg:cg-construction}, we first construct a call graph and collect metadata about classes and methods. Given the classpath \(CP\), a target class \(C_t\), and a target method name \(m_t\), the procedure consists of three phases.

In the first phase, we load all bytecode from the classpath and extract the set of classes \(\mathcal{C}\) (lines 1--3). This phase establishes the scope of the global analysis required for subsequent call graph construction.

In the second phase, we collect a set of metadata maps, including the class visibility map \(\textit{classVis}\), the method visibility map \(\textit{methodVis}\), the constructor map \(\textit{ctors}\), and the implementation map \(\textit{impls}\). The visibility maps record whether each class or method is publicly accessible, which is essential for identifying valid invocation entry points and feasible call paths. The constructor map stores public constructors for each class, enabling object instantiation during test generation. The implementation map maps supertypes (including superclasses and interfaces) to their public concrete implementations, facilitating the resolution of virtual and interface dispatch during call graph construction.

In the third phase, we adopt Class Hierarchy Analysis (CHA)~\citep{dean1995optimization} to construct the call graph \(CG\), as it provides a favorable balance between scalability and recall. Since the goal of \tech{} is to provide the LLM with candidate invocation paths that may reach the target method, we prioritize broad coverage of potential call relationships over aggressive pruning of infeasible edges. Compared with more precise yet computationally expensive alternatives, such as RTA~\citep{bacon1996fast} or points-to-based analyses~\citep{smaragdakis2015pointer}, CHA enables lightweight whole-project analysis without requiring complete entry-point information, making it particularly suitable for project-level iterative test generation.

\added{For CHA construction, we use as the root set \(\mathcal{R}\) (line~19) only the public methods of all classes on the classpath (i.e., from the project and its libraries), rather than all methods. This choice matches the practical test-generation setting, where only externally accessible methods can serve as entry points. It also reduces spurious call paths that start from private helpers, improving both the relevance and scalability of the extracted call chains. Note that \(\mathcal{R}\) only determines where CHA starts: from a public entry point, CHA still traverses non-public callees, so a non-public focal method can be reached through the public methods that call it (directly or via non-public helpers).}

\added{Because the generated test class is placed in the focal class's package, package-private and protected methods in that package could also be used as entry points. We nevertheless restrict \(\mathcal{R}\) to public methods to stay consistent with our filtered call paths and initialization context (which include only public members) and to limit the cost of path enumeration, since each additional entry point adds a starting node and the enumeration cost grows exponentially with path length. We discuss this restriction as a threat to validity.}

As discussed in Section~\ref{sec:motivate}, a single method name may correspond to multiple overloads with different parameter lists. To ensure coverage of all potential invocation paths to the target method, we collect all methods in \(C_t\) whose names match \(m_t\) as the target method set \(\mathcal{T}\) (line 21).

The outputs of this step include the call graph \(CG\), the metadata maps (\(\textit{classVis}\), \(\textit{methodVis}\), \(\textit{ctors}\), and \(\textit{impls}\)), the root set \(\mathcal{R}\), and the target method set \(\mathcal{T}\).

\begin{algorithm}[t]
\caption{Depth-Limited Call Path Extraction}
\label{alg:path-extraction}
\begin{algorithmic}[1]

\REQUIRE Call graph $CG$, entry points $\mathcal{R}$, target methods $\mathcal{T}$, max depth $d_{max}$
\ENSURE Call paths $\mathcal{P}$

\STATE \textbf{Phase 1: Forward Adjacency Construction}
\STATE $F \gets \emptyset$
\FOR{each edge $(m_s, m_t) \in CG.\textit{edges}$}
    \STATE $F[m_s].\textsc{Add}(m_t)$
\ENDFOR

\STATE \textbf{Phase 2: Backward Reachability Analysis}
\STATE $\textit{canReach} \gets \mathcal{T}$
\REPEAT
    \FOR{each $(m, \textit{callees}) \in F$}
        \IF{$m \notin \textit{canReach}$ and $\textit{callees} \cap \textit{canReach} \neq \emptyset$}
            \STATE $\textit{canReach}.\textsc{Add}(m)$
        \ENDIF
    \ENDFOR
\UNTIL{no changes}

\STATE \textbf{Phase 3: Depth-Limited DFS from Reachable Entry Points}
\STATE $\mathcal{P} \gets \emptyset$
\FOR{each $r \in \mathcal{R} \cap \textit{canReach}$}
    \STATE \textsc{DFS}($F$, $r$, $\mathcal{T}$, $d_{max}$, $0$, $[]$, $\emptyset$, $\mathcal{P}$)
\ENDFOR

\RETURN $\mathcal{P}$

\vspace{0.5em}
\hrule
\vspace{0.5em}

\textbf{Procedure} \textsc{DFS}($F$, $m$, $\mathcal{T}$, $d_{max}$, $d$, $S$, $\textit{visited}$, $\mathcal{P}$)
\STATE $S.\textsc{Push}(m)$; $\textit{visited}.\textsc{Add}(m)$
\IF{$m \in \mathcal{T}$ and $d \leq d_{max}$}
    \STATE $\mathcal{P}.\textsc{Add}(\textsc{Copy}(S))$
\ENDIF
\IF{$d < d_{max}$}
    \FOR{each $s \in F[m] \setminus \textit{visited}$}
        \STATE \textsc{DFS}($F$, $s$, $\mathcal{T}$, $d_{max}$, $d{+}1$, $S$, $\textit{visited}$, $\mathcal{P}$)
    \ENDFOR
\ENDIF
\STATE $S.\textsc{Pop}()$; $\textit{visited}.\textsc{Remove}(m)$

\end{algorithmic}
\end{algorithm}

\subsubsection{Depth-Limited Call Path Extraction}
\label{sec:path-extraction}

After constructing the call graph and collecting metadata, we extract candidate call paths from entry points to target methods. Given the call graph \(CG\), the set of entry points \(\mathcal{R}\), the set of target methods \(\mathcal{T}\), and a maximum search depth \(d_{max}\), we perform a depth-limited depth-first search (DFS) to enumerate all feasible call sequences, as shown in Algorithm~\ref{alg:path-extraction}.

The extraction procedure consists of three phases. First, we construct a forward adjacency map \(F\) from \(CG\) to support efficient traversal (lines 1--5). Second, we perform a backward reachability analysis to identify methods that can potentially reach any target method, pruning unreachable nodes and thereby reducing the search space for path enumeration (lines 6--14). Finally, we initiate a depth-limited DFS from each reachable entry point, enumerating all paths that lead to target methods while respecting the maximum depth constraint (lines 15--19). The DFS maintains a stack \(S\) for the current path and a visited set to prevent cycles; whenever a target method is reached within the depth limit, the current path is added to the set of raw call paths \(\mathcal{P}\).

The maximum search depth \(d_{max}\) is introduced to prevent the DFS from exploring exponentially many paths in large projects. Without this limit, the worst-case complexity grows roughly as $O(k_{\text{out}}^{d})$, where \(k_\text{out}\) is the average out-degree of a method and \(d\) is the path length, which is computationally infeasible for project-level analysis~\citep{dean1995optimization,horwitz1990interprocedural}. Importantly, we do not perform extraction and dependency resolution (discussed later) at each generation; instead, we run them once at the beginning as a preprocessing step and reuse the results throughout the process.

The resulting set \(\mathcal{P}\) contains all potential sequences of method invocations from entry points to target methods within the depth bound. These paths may include methods that are not publicly accessible or belong to non-public classes, which are infeasible for test generation. We therefore perform a filtering and initialization inference step in the next phase, as described in Section~\ref{sec:path-filtering}.

\begin{algorithm}[tbp]
\caption{Call Path Filtering and Initialization Set Extraction}
\label{alg:path-filtering}
\begin{algorithmic}[1]

\REQUIRE Call paths $\mathcal{P}$, target class $C_t$, metadata maps
\ENSURE Filtered call paths $\mathcal{P}'$, initialization set $\mathcal{I}$

\STATE \textbf{Phase 1: Visibility-Based Path Filtering}
\STATE $\mathcal{P}' \gets \emptyset$
\FOR{each path $p \in \mathcal{P}$}
    \STATE $p' \gets []$
    \FOR{each method $m \in p$}
        \IF{$\textit{methodVis}[m] = \text{public}$ \AND
            $\textit{classVis}[m.\textit{class}]$ is public \AND
            $m.\textit{class}$ is not abstract}
            \STATE $p'.\textsc{Append}(m)$
        \ENDIF
    \ENDFOR
    \IF{$p' \neq \emptyset$}
        \STATE $\mathcal{P}'.\textsc{Add}(\textsc{Deduplicate}(p'))$
    \ENDIF
\ENDFOR

\STATE \textbf{Phase 2: Class Collection for Initialization}
\STATE $\mathcal{C} \gets \{C_t\}$
\FOR{each path $p \in \mathcal{P}'$}
    \FOR{each method $m \in p$}
        \STATE $\mathcal{C}.\textsc{Add}(m.\textit{declaringClass})$
        \STATE $\mathcal{C}.\textsc{AddAll}(\textsc{GetParamTypes}(m))$
    \ENDFOR
\ENDFOR

\STATE \textbf{Phase 3: Transitive Expansion via Constructors}
\REPEAT
    \STATE $\textit{newClasses} \gets \emptyset$
    \FOR{each $c \in \mathcal{C}$}
        \FOR{each public constructor $\textit{ctor} \in \textit{ctors}[c]$}
            \FOR{each param type $t \in \textsc{GetParamTypes}(\textit{ctor})$}
                \IF{$t \notin \mathcal{C}$ and $t$ is not primitive}
                    \STATE $\textit{newClasses}.\textsc{Add}(t)$
                \ENDIF
            \ENDFOR
        \ENDFOR
    \ENDFOR
    \STATE $\mathcal{C} \gets \mathcal{C} \cup \textit{newClasses}$
\UNTIL{$\textit{newClasses} = \emptyset$}

\STATE \textbf{Phase 4: Initialization Set Construction}
\STATE $\mathcal{I} \gets \emptyset$
\FOR{each class $c \in \mathcal{C}$}
    \FOR{each public constructor $\textit{ctor} \in \textit{ctors}[c]$}
        \STATE $\mathcal{I}.\textsc{Add}(\textit{ctor})$
    \ENDFOR
\ENDFOR

\RETURN $\mathcal{P}'$, $\mathcal{I}$

\end{algorithmic}
\end{algorithm}

\subsubsection{Call Path Filtering and Dependency Resolution}
\label{sec:path-filtering}

Finally, we refine the call paths $\mathcal{P}$ and resolve dependencies based on the extracted paths, as described in Algorithm~\ref{alg:path-filtering}. 
This procedure consists of four phases: the first performs filtering, and the remaining three handle dependency resolution.  

In the filtering phase, call paths extracted in the previous step may include methods that are not publicly accessible or belong to non-public or abstract classes, which are infeasible as entry points for test generation. 
We therefore filter each path by retaining only methods that are public and belong to public, non-abstract classes, yielding the set of \textit{filtered call paths} $\mathcal{P}'$ (lines 1–9). 
Additionally, duplicate method occurrences within each path are removed to eliminate redundant analyses (lines 10–12).  

Based on these filtered call paths, we perform dependency resolution to identify all classes that may need to be initialized for test generation. 
This set includes the target class \(C_t\), the declaring classes of all methods appearing in $\mathcal{P}'$, and the types of all method parameters (lines 14–21). 
To ensure completeness, we further expand this class set transitively by incorporating the parameter types of public constructors, thereby capturing all classes that may be involved in object creation along the filtered call paths (lines 22–35). The process terminates when no new dependent classes are discovered. 

Finally, we construct the \textit{initialization set} $\mathcal{I}$ by collecting all public constructors of the classes identified in the dependency resolution phase (lines 36–42). 
This initialization set, together with the filtered call paths $\mathcal{P}'$, provides the LLM with all necessary information to correctly instantiate objects and generate tests that can reliably reach the target methods.  

The output of this algorithm ($\mathcal{P}'$ and $\mathcal{I}$) serves as input for the test generation phase of \tech{}, ensuring that both feasible call paths and all required object initializations are available to guide high-coverage test generation.

\subsection{Generation Phase} \label{sec:genphase}

After extracting and filtering call-chain information and resolving inter-class dependencies, we integrate these artifacts into the generation prompt to guide the LLM toward producing high-quality tests. As outlined in Section~\ref{sec:overview}, this phase operates within an iterative generate--fix loop that alternates between test generation and test fixing.
Specifically, once the LLM produces candidate tests in this phase, all generated tests are compiled and executed. Failed tests are collected and forwarded to the fixing phase, while successfully passing tests are retained for line- and branch-coverage evaluation. The iterative process continues until one of the stopping criteria is met.
We terminate the generation loop when either (1) the maximum number of generation iterations, \(N_{gen}\), is reached, or (2) no improvement in line or branch coverage is observed for \(m\) consecutive iterations, indicating coverage stagnation and suggesting that further generation is unlikely to yield additional benefits.

The prompting strategy, consisting of both system and user prompts, follows standard LLM prompting conventions~\citep{liu2023pre,ouyang2022training,qwencodedoc} and is explicitly designed to guide the model to instruct the model on how to leverage the provided call-chain and initialization contexts, with particular emphasis on three principles: (1) prioritizing direct invocation of focal methods whenever feasible, (2) enabling indirect invocation through call-chain-guided entry methods when direct calls are insufficient due to the inability of the LLM to synthesize valid test inputs for directly invoking the focal method (e.g., when required inputs are too complex), and (3) ensuring that generated tests are compilable, deterministic, and semantically meaningful.

The system prompt and the main components of the user prompt are illustrated in Figures~\ref{fig:sys_prompt} and~\ref{fig:user_prompt}, respectively. The complete prompts are omitted from the main text due to space limitations and are instead included in the replication package. In Figure~\ref{fig:user_prompt}, we intentionally omit prompt components that are already widely adopted in prior LLM-based test generation approaches~\citep{gu2025llm,zhang2024testbench,wu2025generating,wang2025mutation,xu2025hallucination}, such as generic instructions for including the focal class source code and output-format specifications.

The system prompt design follows established prompt engineering guidelines that advocate placing explicit instructions early, separating task directives from contextual information, and providing precise behavioral constraints and desired output properties. Accordingly, our system prompt, as shown in Figure~\ref{fig:sys_prompt}, first establishes global generation principles, including direct-first invocation, initialization-aware object construction, call-chain-guided indirect reasoning, and assertion-oriented deterministic test generation.

The initialization-first principle (1) is motivated by our empirical observation that initialization routines frequently constitute a substantial portion of executable class behavior and are often essential for reaching deeper program states and execution paths. For example, in \texttt{LUDecomposition} from Math-2f, initialization-related methods account for approximately 30\% of the class code, while \texttt{ArrayFieldVector} exposes 20 public constructors. These observations justify explicitly prioritizing initialization reasoning in the system prompt.

The direct (2) and indirect (3) invocation principles are motivated by complementary coverage scenarios. Direct invocation typically offers the most straightforward route to covering focal methods. However, when the focal method is non-public or when realistic object interactions better reflect intended usage, indirect invocation via earlier public methods in the call-chain becomes more effective. To further improve reasoning precision, the system prompt also enforces call-chain- and context-aware reasoning (4), explicitly encouraging the model to fully exploit provided call-chain structures and treat overloaded methods as distinct coverage targets. As discussed in Section~\ref{sec:motivate}, overloaded methods account for up to 27\% of methods in some repositories, making this distinction practically important. The test quality and stability principle (5) is motivated by the inherent characteristics of automated test generation, where coverage-oriented exploration may produce tests that are syntactically valid but semantically weak or unstable~\citep{schafer2023empirical,fraser2011evosuite}. In particular, generated tests may pass compilation but still suffer from non-determinism (e.g., due to reliance on time, randomness, or environment-dependent states) or lack meaningful assertions.

The user prompt, shown in Figure~\ref{fig:user_prompt}, complements the system prompt by providing structured, task-specific context. In particular, it provides detailed usage instructions for both the call-chain and initialization contexts, along with concise examples that demonstrate how to apply direct and indirect invocation strategies (lines 13--17) and how to construct valid receiver objects from initialization metadata (lines 21--23). Line~24 instructs the LLM to use this metadata, and we replace lines 18 and 25 with the extracted call-chain and initialization contexts, respectively. In addition, we include source files of related classes (lines 26-27) to expose dependency semantics and API behaviors that may be critical for generating meaningful tests.

A key practical challenge is prompt length. For classes with rich dependencies or long transitive call chains, the resulting call-chain context, initialization metadata, and related source files can become prohibitively large. To ensure the prompt remains within the LLM token budget, we bound both the call-chain exploration depth in Algorithm~\ref{alg:path-extraction} and the number of top-frequency related source files by $d_{max}$. As discussed in Section~\ref{sec:implement}, these parameters were determined through a preliminary study. This design also aligns with the exponential worst-case complexity of DFS-based call-chain extraction discussed in Section~\ref{sec:path-extraction}. In particular, a larger $d_{max}$ may lead to an exponential increase in the number of explored call-chain combinations, substantially prolonging extraction time and, in extreme cases, making the analysis practically difficult to terminate within reasonable resource limits.

\begin{figure}[tbp]
\centering
\begin{minipage}{0.95\linewidth}
\begin{tcolorbox}[title=The System Prompt for Test Generation,
    fonttitle=\bfseries\color{black},
colback=gray!8,colframe=gray!40,boxrule=0.5pt]
\small
You are an expert Java software tester. Your task is to generate high-quality and reliable test cases that maximize coverage of focal methods in a target class, using the provided call-chain and dependency information. Follow these principles:\\
1. Initialization-first testing:\\
   - Prioritize direct tests for constructors and initialization methods.\\
   - Explore diverse inputs to cover branching initialization logic and state setup.\\
   - Use lightweight mocking only when dependencies cannot be naturally resolved.\\
2. Direct invocation priority:\\
   - Prefer directly invoking focal methods and initializers.\\
   - Use varied and boundary inputs to maximize branch coverage.\\
3. Indirect invocation when necessary:\\
   - Use call-chain-based indirect invocation only when direct calls are insufficient.\\
   - Respect the provided call-chain structure.\\
4. Call-chain- and context-aware reasoning:\\
   - Exercise all provided call-chain information when applicable.\\
   - Treat overloaded methods as distinct targets when applicable.\\
5. Test quality and stability:\\
   - Produce compilable, deterministic, and meaningful tests.\\
   - Include assertions for observable behavior whenever possible.
\end{tcolorbox}
\end{minipage}
\caption{The system prompt used in \tech{}}
\label{fig:sys_prompt}
\end{figure}

\begin{figure}[tbp]
\centering
\begin{minipage}{0.95\linewidth}
\begin{tcblisting}{
title=Main Excerpt of the User Prompt for Test Generation,
fonttitle=\bfseries\color{black},
colback=gray!8,
colframe=gray!40,
boxrule=0.5pt,
listing only,
listing options={
    basicstyle=\small\rmfamily,
    breaklines=true,
    breakatwhitespace=false,
    columns=fullflexible,
    keepspaces=true,
    numbers=left,
    numbersep=8pt,
    numberstyle=\scriptsize\color{gray},
    showstringspaces=false,
    language={},
    breakindent=0pt,
    escapechar=§
}}
## §\textbf{Call-Chain Context (Test Design Driver)}§
You are provided with static call-chain structures, each terminating at the focal method, i.e., the focal method is the final node in every call-chain. These structures represent candidate execution routes that may be exercised through appropriate input construction, object initialization, or method invocation sequences.
A focal method may be covered either by:
(1) directly invoking the focal method itself, or
(2) indirectly invoking any earlier public method within a call-chain that eventually reaches the focal method.
For each call-chain structure, the focal method always appears as the final item, while preceding nodes denote methods that may transitively invoke it. To achieve indirect coverage, any earlier public method in the call-chain may be selected as the entry point. Once selected, initialize the corresponding class using the Initialization section below, then invoke the chosen method so that execution follows the call-chain toward the focal method.
If the focal method itself is non-public, coverage should be achieved through indirect invocation via accessible public methods within the call-chain.
### §\textbf{Call-Chain Context}§
The following example first illustrates two possible invocation strategies, namely direct and indirect coverage. It also demonstrates how call-chain information can be used to select an appropriate public entry point. The complete call-chain context is then provided below.
For direct coverage, the focal method can be invoked directly if it is publicly accessible;
For indirect coverage, an earlier public method in the call-chain may be selected as the entry point, e.g., MapSerializer#withFilterId(Object), whose returned MapSerializer object can be used to transitively reach ClassUtil#verifyMustOverride(...).
Prefer indirect invocation when the focal method is non-public or when execution better exercises realistic object interactions.
§\textbf{\textit{Example:}}§
    Context 1: MapSerializer#withFilterId(Object): MapSerializer (public)
                    -> ClassUtil#verifyMustOverride(...)
    Context 2: BeanDeserializerFactory#withConfig(...) (public)
                    -> ClassUtil#verifyMustOverride(...)
§\textit{\{\{ CALL\_CHAINS \}\}}§
### §\textbf{Initialization Context}§
The following example first illustrates how initialization information should be used to construct valid objects. The actual initialization context for all relevant classes is then provided below, together with their access modifiers.
§\textbf{\textit{Example:}}§
    com.fasterxml.jackson.databind.util.ClassUtil (public final): ClassUtil()
This indicates that the class ClassUtil is publicly accessible and can be directly instantiated using its public constructor ClassUtil().
Use this information to construct valid objects, select accessible entry methods, and satisfy dependency requirements during test generation.
§\textit{\{\{ INITIALIZATION\_INFO \}\}}§
### §\textbf{Source Files of Related Classes}§
§\textit{\{\{ RELATED\_SOURCE\_FILES \}\}}§
\end{tcblisting}
\end{minipage}
\caption{Main excerpt of the user prompt used in \tech{}}
\label{fig:user_prompt}
\end{figure}

\subsection{Fixing Phase} \label{sec:fixphase}

After each generation iteration, once the LLM produces candidate tests, we compile and execute all generated tests. Failed tests are collected and forwarded to the fixing phase, where execution feedback is leveraged to repair non-compilable or failing tests. Specifically, we incorporate failed tests together with their corresponding error messages into the fixing prompt, allowing up to \(N_{fix}\) repair attempts within each generation iteration. \added{After each repair attempt, the repaired tests are recompiled and re-executed. Tests that now pass are retained and contribute to coverage, whereas the remaining failures---together with their updated error messages---are forwarded to the next repair attempt (the ``Fix all?'' check in Figure~\ref{fig:overview}). Tests that still fail after \(N_{fix}\) attempts are discarded.}

The fixing prompt is illustrated in Figure~\ref{fig:fixing_prompt}.

The fixing prompt explicitly instructs the model to focus solely on repairing issues in the test code while preserving the semantics of the production code. Typical repair targets include compilation errors (e.g., missing imports and incorrect API usage leading to type mismatches) and runtime failures (e.g., invalid assertions and incorrect object construction), which are among the most common failure modes in LLM-generated tests. \added{Here, we define an invalid assertion as one whose expected value does not match the output of the current implementation, which the fixing prompt treats as the ground truth (Figure~\ref{fig:fixing_prompt}).}

Prior studies show that the effectiveness of LLM-based test repair depends strongly on the model's ability to correctly interpret compiler and runtime feedback, with reported repair rates typically around 50\%. Therefore, we do not expect all failing tests to be successfully repaired. Instead, our goal is to recover a substantial portion of failing tests (our approach fixes around 53\%) so that they can still contribute to cumulative coverage improvement in subsequent generation iterations. This design also prevents the overall iterative process from being prematurely stalled by failed fixing, especially when no successful repairs are observed for \(m\) consecutive iterations.

To keep the fixing prompt within the model's token budget, we include all failed tests from the current generation but retain only the most informative parts of the associated error feedback: error line numbers and concise failure reasons. In addition, we restrict the provided source files to the top-$d_{max}$ most frequently occurring classes in the extracted call-chain context, consistent with Section~\ref{sec:genphase}.

\begin{figure}[tbp]
\centering
\begin{minipage}{0.95\linewidth}
\begin{tcolorbox}[
title=Fixing Prompt for LLM-based Test Repair,
fonttitle=\bfseries\color{black},
colback=gray!8,
colframe=gray!40,
boxrule=0.5pt,
enhanced,
parbox=true
]
\small
\textbf{\#\# System Prompt for Test Fixing}\\
You are a program repair assistant specialized in fixing failing unit tests.\\
Your goal is to modify the generated test so that it is syntactically correct, compilable, and consistent with the system behavior implied by the source code.\\
You should NOT modify the production code.\\
You should ONLY fix issues in the test code, including: API usage errors, invalid assertions, missing imports, and incorrect object construction.\\
Maintain the original testing intent as much as possible.\\
\textbf{\#\# User Prompt for Test Fixing}\\
You are given a failing or non-compilable test generated by an LLM-based test generation system, along with relevant source files and optional error messages.\\
Your task is to fix the test so that it compiles and behaves correctly according to the source code semantics.\\
\textbf{\#\# Failing Tests and Error Messages}\\
\textit{{{FAILING\_TESTS\_AND\_ERRORS}}}\\
\textbf{\#\# Source Files of Related Classes}\\
The following source files define the ground-truth behavior and APIs used by the test.\\
Use them as the primary reference for fixing the test:\\
\textit{{{RELATED\_SOURCE\_FILES}}}

\end{tcolorbox}

\end{minipage}
\caption{Fixing prompt used for LLM-based test repair with source-code grounding}
\label{fig:fixing_prompt}
\end{figure}

\section{Evaluation} \label{sec:eval}
Our approach (\tech{}) incorporates call-chain information to explicitly model external dependencies in LLM-based project-level test generation. As discussed earlier, \panta{} represents a recent, state-of-the-art LLM-based project-level testing framework that iteratively provides uncovered execution paths of the focal class as feedback to guide test generation toward higher coverage. However, \panta{} does not explicitly capture cross-module dependencies, resulting in degraded performance for classes with rich external dependencies.

In our evaluation, we compare \tech{} with the original \panta{} on the same benchmark, Defects4J. 
Unlike \tech{}, which generates tests for each focal method individually, \panta{} generates tests at the focal-class level, producing tests for all focal methods within a class in a single run. As a result, even though \tech{} uses fewer iterations per focal method than \panta{}, the total number of generation iterations can be substantially larger when aggregated at the class level. For example, while \panta{} may perform 30 iterations for a single focal class, a class containing 20 focal methods would require \tech{} to perform 200 iterations if tests for each method are generated with 10 iterations. This naturally leads to a higher overall time cost.
To ensure a fair comparison, we also include a time-bounded variant of \panta{}, denoted \pantavariant{}, constrained by the same time budget as \tech{}. To quantify the contribution of static analysis in \tech{}, we also report the performance of \techvariant{}, whose prompts exclude the execution context extracted in Section~\ref{sec:gen}.


Since modern LLMs are trained on large-scale public code repositories, Defects4J projects may partially overlap with their training corpora. This overlap could inflate results due to data leakage or memorization. Although both \tech{} and the baseline (originally evaluated on Defects4J) are equally affected by this issue, we also evaluate them on projects not included in the LLM’s training data. This ensures our results reflect realistic performance on truly unseen code and better demonstrate the practical utility of our approach.
Moreover, these additional projects are structurally more complex than those in Defects4J. As discussed in Section~\ref{sec:subjects}, the Defects4J projects are organized as single-module repositories, where classes are primarily located within a single source directory, and dependencies are mainly inter-class. In contrast, the new projects are multi-module repositories involving multiple source directories and more complex dependencies across both classes and modules. Evaluating on these projects therefore not only reduces the risk of training data leakage but also provides stronger evidence of \tech{}'s generalization ability in more realistic, challenging repository-level testing scenarios.

To summarize, our evaluation is designed to answer the following research questions:

\begin{itemize}
    \item \textbf{RQ1:} How does CAT compare with existing test generation approaches in terms of effectiveness, efficiency, and generated test characteristics?
    \item \textbf{RQ2:} To what extent do call-chain awareness and dependency context contribute to CAT’s effectiveness, particularly through indirect invocations?
    \item \textbf{RQ3:} How does \tech{} perform on projects that are not included in the training data of the underlying LLM, and how does it compare with existing approaches?
\end{itemize}
RQ1 compares overall effectiveness and efficiency with \panta{} and its time-bounded variant \pantavariant{}, and compares effectiveness with EvoSuite. RQ2 investigates the contribution of call-chain and dependency contexts through ablation analysis. RQ3 examines \tech{}'s generalization ability on unseen and more complex multi-module projects.

\subsection{Experiment Setup}
\subsubsection{Subjects}\label{sec:subjects}
We reuse the projects from the original \panta{} experiments. Table~\ref{tab:project_info} presents the statistics of the 14 projects used in our evaluation. Following \panta{}'s experiments, we extract focal methods from these 14 projects. To focus on complex code, rather than generating tests for all classes, we select non-abstract public classes that contain at least one method with cyclomatic complexity greater than 10. The \textit{Project} column lists the project name and its Defects4J version number. The \textit{\#Class} column indicates the number of focal classes, and \textit{\#MUTs} reports the total number of focal methods in the selected classes for each project. The CYC$_{max}$ column reports the maximum cyclomatic complexity among the selected methods under test. Additionally, \panta{} uses this value as the maximum number of iterations. Overall, our evaluation of Defects4J projects covers 130 classes comprising 2,971 MUTs. \added{We use the list of focal classes published with \panta{}~\citep{gu2025llm}; the MUTs are the non-private methods of these classes, excluding trivial getters, setters, and empty methods.}

\begin{table}[tbp]
\centering
\small
\caption{Information on the Defects4J projects.}
\label{tab:project_info}
\begin{tabular}{l l c c c}
\toprule
\textbf{Identifier} & \textbf{Project} & \textbf{\#Class} & \textbf{\#MUTs} & \textbf{CYC$_{max}$} \\
\midrule
JacksonXml & JacksonXml-5f & 4 & 118 & 29\\
Csv & Csv-16f & 3 & 72 & 25\\
Collections & Collections-28f & 5 & 218 & 32\\
Gson & Gson-16f & 4 & 75 & 37\\
Cli & Cli-40f & 2 & 31 & 11\\
JacksonCore & JacksonCore-26f & 9 & 161 & 30 \\
JxPath & JxPath-22f & 12 & 115 & 32\\
Jsoup & Jsoup-93f & 8 & 171 & 22\\
Codec & Codec-18f & 7 & 78 & 31 \\
Compress & Compress-47f & 9 & 65 & 22\\
JacksonDatabind & JacksonDatabind-112f & 9 & 371 & 22\\
Time & Time-13f & 11 & 458 & 30\\
Lang & Lang-4f & 17 & 586 & 30\\
Math & Math-2f & 30 & 452 & 31\\
\midrule
\textbf{Total} &  & 130 & 2,971 & - \\
\bottomrule
\end{tabular}
\end{table}

Additionally, to better evaluate the practical performance of \tech{} and \panta{} on projects unseen by the LLM (Qwen3-Coder-30B, described later), we collect four other new Java projects from GitHub. Specifically, we search for “Java projects,” select repositories released after April 2025 (i.e., after the LLM's release date) with more than 500 stars. After excluding projects that cannot be successfully built with Maven, four projects remain. For these projects, we directly clone the corresponding GitHub repositories.

Unlike the Defects4J subjects, where each project typically corresponds to a single source root, these GitHub projects commonly adopt \textit{multi-module Maven structures} comprising multiple source folders and interdependent modules. This setting introduces more realistic project-level dependency relationships and cross-module call chains, making it particularly suitable for evaluating \tech{}'s robustness in large-scale, practical settings.

We then select focal methods using the same criteria as in the Defects4J evaluation. One exception is the Binance project, which contains a large number of classes (377 classes after initial filtering). To keep the evaluation manageable, we further exclude classes that contain no methods with cyclomatic complexity greater than 20, resulting in 41 classes.

As in Table~\ref{tab:project_info}, Table~\ref{tab:project_info_self} reports statistics for these LLM-unseen projects used in our evaluation. In total, 112 classes containing 2,707 MUTs are included.

\begin{table}[tbp]
\centering
\small
\caption{Information on the unseen projects.}
\label{tab:project_info_self}
\begin{tabular}{l l l c c c}
\toprule
\textbf{Identifier} & \textbf{Project} & \textbf{Version Tag} & \textbf{\#Class} & \textbf{\#MUTs} & \textbf{CYC$_{max}$}\\
\midrule
Adj & adk-java~\citep{adj} & 0058300b & 19 & 86 & 36\\
Astron & astron-agent~\citep{astron} & 5e7a0e61 & 39 & 259 & 38\\
Binance & binance-connector-java~\citep{binance} & f4c198cb & 41 & 2,335 & 29\\
Joyagent & joyagent-jdgenie~\citep{joyagent} & 7142f415 & 13 & 27 & 30\\
\midrule
\textbf{Total} & - & - & 112 & 2,707 & - \\
\bottomrule
\end{tabular}
\end{table}

\subsubsection{Process} \label{sec:process}
\added{The static analysis of \tech{} is performed offline as a one-time preprocessing step before test generation. For each focal method, we extract its call-chain and dependency contexts as described in Section~\ref{sec:gen}. This preprocessing takes 8.35 hours for the 2,971 focal methods in Defects4J and 61.63 hours for the 2,707 focal methods in the unseen projects. Once \tech{} extracts contexts for all focal methods, it proceeds to the LLM-based generation and fixing phases, reusing the precomputed contexts across iterations without repeating the static analysis. We therefore report this preprocessing cost separately rather than including it in the end-to-end test generation time.}

To address RQ1, we first run the original \panta{} with its default configuration on the Defects4J dataset using the provided scripts, selecting 14 projects for evaluation. \added{As in the evaluation of \panta{}~\citep{gu2025llm}, tests are generated without existing tests: the test class of each focal class is removed before generation.}
Next, we conducted a preliminary experiment on a random sample of 20 focal methods from all projects. We observed that after 10 iterations, \tech{} coverage improvement becomes marginal. Accordingly, we set the number of iterations for \tech{} to 10 ($N_{gen}=10$) for all focal methods across all projects. We apply the same iteration setting to \techwithpath{}, our variant that augments \tech{} with the uncovered-path information used in \panta{}, to ensure comparability.
Notably, \panta{} generates tests for all focal methods within a class in a single run, using the maximum cyclomatic complexity as the number of iterations, which is typically much higher than 10. As shown in Table~\ref{tab:project_info}, the average maximum cyclomatic complexity (CYC$_{max}$) for the selected methods under test in the Defects4J projects is 27, ranging from 11 to 37.
To ensure a fair comparison, we also run \panta{} with a maximum of 100 iterations and impose a class-specific time budget equal to the time \tech{} consumes on the corresponding focal class. We terminate \panta{} once this time budget is reached, yielding a time-constrained variant denoted \pantavariant{}. As an additional exploratory analysis, we further evaluate \techwithpath{} under the same iteration and time budget settings as \tech{} to examine whether combining uncovered-path information with call-chain and dependency context yields further gains.
\added{We report line and branch coverage measured by JaCoCo~\citep{jacoco}, as well as mutation scores computed by PITest~\citep{pitest2025}, to assess defect-detection effectiveness, following established practices in prior work~\citep{wang2025mutation,xu2025hallucination,harman2025mutation}. Additionally, we use PITest with its DEFAULTS mutation operator group to measure each test suite's mutation score. The DEFAULTS group comprises mutation operators commonly adopted in the literature~\cite{fraser2015achieving, wang2026comprehensive}. According to the official PITest documentation, these operators are designed to generate mutants that are relatively difficult for tests to detect while limiting the generation of equivalent mutants.}
\added{All metrics are computed on each approach’s final test suite, comprising the tests retained after all generation and repair iterations. Following \panta{}, we compute these metrics at the focal-class level. For each focal class, we combine all retained tests for its focal methods and obtain line and branch coverage from the corresponding JaCoCo reports. Mutation scores are computed using PITest, with the focal classes specified as mutation targets. The value reported for a project is the mean over its focal classes. We apply the same measurement procedure to \tech{}, its variants, and all baselines. }

\added{In addition to coverage and mutation score, we report the total number of LLM tokens consumed, end-to-end test generation time, the number of retained test methods, and the proportion of retained test methods that use a mocking framework. The total token consumption includes both input and output tokens across all LLM calls made during test generation and repair. End-to-end time measures the wall-clock time from the first LLM invocation to the final test suite, including LLM inference, test repair, compilation, and test execution. It excludes the static-analysis phase used to construct call-chain and dependency contexts, since this analysis runs offline as a one-time preprocessing step and its results can be reused throughout test generation. The number of retained tests is the number of test methods remaining in the final test suite after all generation and repair iterations. \tech{} retains every test that compiles and passes, including tests that do not increase coverage, and does not remove duplicated tests; \panta{} likewise adds every passing test to its test file~\citep{gu2025llm}. The reported number of retained tests may therefore include duplicated tests, which we did not measure; since a duplicated test executes the same code and checks the same behavior as the test it duplicates, it can inflate the test count but not coverage or the mutation score. To measure the use of mocking, we inspect each retained test method and classify it as using mocking if the method itself, or its test setup required for execution, invokes APIs from the adopted mocking framework (e.g., Mockito). We then compute the mocking rate as the number of retained test methods using mocking divided by the total number of retained test methods. We apply the same metrics and measurement procedures to RQ3. To further analyze API hallucinations, we examined the compilation diagnostics and test execution results of all tests generated by \tech{} on the Defects4J projects, checked the referenced program elements against the project source code, classified the identified hallucinations into two main categories reported in detailed results, and calculated the percentage of all generated tests affected by each category.}

\added{To address RQ2, we construct three variants of \tech{}, denoted as \techvariant{}, \techchain{}, and \techdep{}, by disabling both call-chain awareness and dependency resolution, retaining only call-chain awareness, and retaining only dependency resolution, respectively, as described in Section~\ref{sec:gen}. Specifically, we remove the corresponding contextual information from both the generation and fixing prompts described in Sections~\ref{sec:genphase} and~\ref{sec:fixphase}. We evaluate these variants on Defects4J using the same configuration and evaluation metrics as \tech{} in RQ1, allowing us to isolate the individual and combined contributions of call-chain awareness and dependency resolution. We further examine the role of indirect invocations enabled by call-chain awareness. Specifically, we quantify the proportion of retained tests that invoke their focal methods indirectly through the extracted call chains and measure the coverage and mutation-score improvements attributable to these tests.}

To address RQ3, we first collect metadata (i.e., class complexity and focal methods within each class) for the four unseen projects described in Section~\ref{sec:subjects}. Since these projects are multi-module and differ from Defects4J, we adapt the evaluation scripts to support multi-module project structures. We then evaluate \tech{} on the selected target classes. For comparison, we also run \panta{} and \pantavariant{} under the same settings.

\added{In RQ1 and RQ3, we additionally compare against EvoSuite using a strong configuration informed by prior studies. Specifically, we use DynaMOSA as the search algorithm, following its empirical evaluation in prior work~\citep{campos2017empirical} and its adoption in subsequent EvoSuite-based studies~\citep{panichella2022test}. We further enable private-API access through reflection and functional mocking, which have been shown to facilitate constructing program states that are difficult to reach through public APIs alone~\citep{arcuri2017private}, and we use a mutation-based strategy for assertion generation. We allocate a generous search budget of 900 seconds per run and focal class, substantially exceeding EvoSuite’s default search budget of 60 seconds and the budgets commonly adopted in prior studies~\citep{campos2017empirical,panichella2022test}. EvoSuite~1.2.0 is executed under the same Java/JDK environment as the other approaches, while the remaining parameters retain their default settings. Since its search is randomized, we run it three times per focal class with different random seeds and report the average line coverage, branch coverage, and mutation score over the three runs. We do not include EvoSuite in the comparison of LLM token consumption, end-to-end generation time, mocking rate, or the number of retained tests. Token consumption is not applicable because EvoSuite does not rely on LLM inference. Its generation time is determined by a fixed search budget and is therefore not directly comparable to the iterative LLM-based generation process; moreover, we intentionally provide EvoSuite with a substantially larger budget of 900 seconds to avoid disadvantaging it. Similarly, EvoSuite incorporates functional mocking as an intrinsic test-generation mechanism, making its use of mocks conceptually different from mocks generated by the LLM-based approaches. We therefore use EvoSuite primarily as a reference baseline for comparing the effectiveness of the generated test suites: line coverage, branch coverage, and mutation score.}

\added{When comparing coverage and mutation scores, we report differences in percentage points (pp) and give the relative improvement over the baseline, $(a-b)/b$, in parentheses.}
To further assess the statistical significance of the effectiveness improvements achieved by \tech{}, we conducted nonparametric analyses in RQ1 and RQ3. Specifically, we used Vargha and Delaney's $A_{12}$ statistic~\cite{arcuri2011practical,vargha2000critique} to measure the effect size based on the line coverage, branch coverage, and mutation score results. The $A_{12}$ statistic ranges from 0 to 1, with 0.5 indicating no difference between the compared approaches. A value greater than 0.5 indicates that \tech{} is more likely to outperform the baseline on the corresponding dataset, while a value below 0.5 suggests the opposite.
In addition, we applied the paired-sample Wilcoxon signed-rank test to line coverage, branch coverage, and mutation score to assess whether \tech{}'s observed improvements are statistically significant. We used a significance level of $\alpha = 0.05$. Therefore, a $p$-value smaller than 0.05 indicates a statistically significant improvement. 

\subsubsection{Implementation} \label{sec:implement}
We implement the dedicated static analysis framework for call-chain extraction and dependency resolution described in Section~\ref{sec:gen} in Java, based on SootUp V2.0.0~\citep{sootup_github}, with approximately 1,200 lines of code. 

For the evaluation pipeline, our implementation initially builds on \panta{}~\citep{panta2025} and its associated scripts for the Defects4J experiments. However, to support the additional GitHub subjects, which commonly adopt multi-module Maven structures as discussed in Section~\ref{sec:subjects}, we substantially extend and refactor the original evaluation framework. In particular, we redesigned project parsing, built the orchestration, and implemented coverage-collection logic to correctly handle multiple modules and cross-module dependencies. Overall, our evaluation framework is therefore only partially based on \panta{} and includes significant extensions to support realistic project-level settings.

Regarding the configuration of \tech{}, we set the maximum number of generation iterations \(N_{gen}\) to 10 for all focal methods across all projects, as described in Section~\ref{sec:process}. We set the maximum call-path search depth \(d_{max}\) to 3 based on the same preliminary experiment. A larger depth substantially increases path-extraction cost, whose worst-case complexity grows exponentially with the average method out-degree and path length, leading to unacceptable analysis overhead, as discussed in Section~\ref{sec:gen}. Finally, we set both \(m\) and \(N_{fix}\) to 3, following the default configuration of \panta{}.

LLMs have rapidly evolved, encompassing both general-purpose models and those specifically tailored for code-related tasks. In this evaluation, we use Qwen3-Coder, which, when our work began, achieved state-of-the-art performance among open-source models on widely used code generation benchmarks, including SWE-Bench~\citep{jimenez2023swe} and LiveCodeBench~\citep{jain2024livecodebench}.
We deploy the Qwen3-Coder 30B model locally using Ollama~\citep{qwen3coder2025} with its default configuration. To ensure a fair comparison, we set the temperature to 0.2 and use JUnit 4 across all experimental settings on Defects4J, consistent with our baseline \panta{}. \added{We also use the same maximum output length of 16,384 tokens per LLM response for all LLM-based approaches, with no explicit limit on the number of tests generated per response.} For the unseen projects, both \panta{} and \tech{} use JUnit 5, following the Maven configurations specified in each project.

We conduct the evaluation on a Linux server equipped with a 56-core CPU, 125 GB of RAM, a single NVIDIA RTX 6000 Ada GPU, and running Ubuntu Linux 24.04.1 LTS. 
Our tool and benchmarks can be found at: \href{https://github.com/Amocy-Wang/CAT}{https://github.com/Amocy-Wang/CAT}.

\subsection{Results and analysis}

\subsubsection{Comparison between \tech{} and baselines.}\label{sec:rq1}

Table~\ref{tab:coverage} compares coverage and mutation score results for our approach \tech{}, \pantavariant{}, \panta{}, and EvoSuite across the Defects4J projects. Bold values indicate the best performance for each project. As described in Section~\ref{sec:process}, \panta{} generates tests for all focal methods of a class in a single run with the maximum number of iterations shown in the last column of Table~\ref{tab:project_info}, whereas \pantavariant{} runs under the same time budget as \tech{}.

\begin{table}[tbp]
\centering
\scriptsize
\setlength{\tabcolsep}{3.5pt}
\renewcommand{\arraystretch}{1.1}
\caption{Comparison of line coverage, branch coverage, and mutation scores across Defects4J projects. Bold values indicate the best performance for each project.}
\label{tab:coverage}
\begin{tabular}{l >{\columncolor{LightGray}}c >{\columncolor{LightGray}}c >{\columncolor{LightGray}}c c c c c c c c c c}
\toprule
\multirow{2}{*}{\textbf{Project}} & \multicolumn{3}{c}{\textbf{\tech{}}} & \multicolumn{3}{c}{\textbf{\pantavariant{}}} & \multicolumn{3}{c}{\textbf{\panta{}}} & \multicolumn{3}{c}{\textbf{EvoSuite}} \\
\cmidrule(lr){2-4} \cmidrule(lr){5-7} \cmidrule(lr){8-10} \cmidrule(lr){11-13}
 & \textbf{Line} & \textbf{Branch} & \textbf{Mut.} & \textbf{Line} & \textbf{Branch} & \textbf{Mut.} & \textbf{Line} & \textbf{Branch} & \textbf{Mut.} & \textbf{Line} & \textbf{Branch} & \textbf{Mut.} \\
\midrule
JacksonXml & 64.16 & 58.34 & \textbf{61.85} & \textbf{64.83} & 54.14 & 37.13 & 57.18 & 52.45 & 27.13 & 63.22 & \textbf{64.50} & 52.41 \\
Csv & \textbf{82.60} & \textbf{70.83} & \textbf{79.60} & 77.28 & 62.11 & 70.63 & 72.57 & 50.81 & 69.23 & 30.10 & 29.17 & 20.31 \\
Collections & \textbf{92.90} & \textbf{90.05} & \textbf{86.28} & 69.79 & 64.74 & 64.39 & 67.72 & 61.82 & 60.66 & 77.45 & 84.49 & 67.55 \\
Gson & \textbf{87.73} & \textbf{75.07} & \textbf{68.87} & 76.78 & 60.41 & 57.29 & 72.35 & 59.10 & 63.93 & 78.93 & 72.12 & 51.70 \\
Cli & \textbf{87.98} & \textbf{78.47} & \textbf{81.81} & 86.28 & 72.32 & 51.56 & 83.21 & 70.91 & 59.08 & 84.10 & 75.66 & 52.14 \\
JacksonCore & \textbf{71.78} & 61.69 & \textbf{58.27} & 63.47 & 54.62 & 43.22 & 55.50 & 49.43 & 41.91 & 66.69 & \textbf{66.53} & 49.91 \\
JxPath & \textbf{74.13} & \textbf{66.26} & \textbf{53.08} & 56.86 & 51.97 & 50.84 & 56.97 & 53.88 & 48.08 & 64.28 & 61.47 & 52.38 \\
Jsoup & \textbf{88.45} & 71.90 & 63.33 & 84.75 & 70.55 & 53.65 & 81.70 & 62.00 & \textbf{64.17} & 82.55 & \textbf{75.29} & 55.62 \\
Codec & \textbf{86.28} & 77.20 & \textbf{59.64} & 82.29 & \textbf{81.19} & 50.28 & 81.00 & 79.70 & 49.12 & 84.27 & 72.11 & 55.06 \\
Compress & \textbf{64.22} & \textbf{59.52} & \textbf{51.86} & 61.55 & 51.80 & 30.25 & 58.74 & 51.49 & 34.89 & 58.52 & 54.63 & 45.62 \\
JacksonDatabind & \textbf{84.11} & \textbf{75.91} & \textbf{83.07} & 62.07 & 53.40 & 60.03 & 56.07 & 47.22 & 50.72 & 73.50 & 67.60 & 59.20 \\
Time & \textbf{92.28} & 84.36 & \textbf{74.47} & 87.33 & 80.89 & 69.06 & 85.33 & 73.69 & 68.94 & 86.12 & \textbf{85.27} & 54.76 \\
Lang & \textbf{86.48} & 78.25 & \textbf{71.75} & 74.35 & 68.33 & 60.14 & 71.64 & 67.99 & 61.01 & 85.46 & \textbf{81.93} & 63.63 \\
Math & \textbf{85.10} & \textbf{77.34} & \textbf{55.31} & 75.38 & 64.20 & 44.09 & 72.82 & 63.11 & 48.92 & 77.52 & 76.06 & 49.15 \\
\midrule
\textbf{Average} & \textbf{82.01} & \textbf{73.23} & \textbf{67.80} & 73.07 & 63.62 & 53.04 & 69.49 & 60.26 & 53.41 & 72.34 & 69.06 & 52.10 \\
\bottomrule
\end{tabular}
\end{table}

On average, \tech{} surpasses \pantavariant{} and \panta{} by 8.94 and 12.52 pp (12.23\% and 18.02\%) in line coverage, and by 9.61 and 12.97 pp (15.11\% and 21.52\%) in branch coverage, respectively. Across most projects, \tech{} outperforms both baselines, highlighting the effectiveness of incorporating call-chain extraction for call context and dependency resolution for dependency context in prompts for project-level test generation. Two exceptions appear: \textit{JacksonXml}, where \tech{} does not achieve the highest line coverage, and \textit{Codec}, where the same holds for branch coverage.

\added{Compared with EvoSuite, \tech{} achieves higher line coverage on all 14 projects, by 9.67 pp (13.37\%) on average, whereas its average branch coverage is only 4.17 pp (6.04\%) higher, with EvoSuite achieving the highest branch coverage on five projects (JacksonXml, JacksonCore, Jsoup, Time, and Lang). This is consistent with EvoSuite's search, which directly targets individual branches as coverage goals.}

For the \textit{JacksonXml} project, \tech{} performs slightly worse than \pantavariant{} by 0.67 pp (1.03\%) in line coverage, while achieving 4.20 pp (7.76\%) higher branch coverage. This suggests that although the richer call-chain and dependency context provided by \tech{} is particularly effective for exploring diverse execution branches, the uncovered-path information used by \pantavariant{} still offers valuable guidance on input constraints and path-specific conditions, which can help cover certain lines of code. In other words, while \tech{} improves overall branch exploration by addressing dependency issues through call-chain graph information in the prompts, challenges related to precise input construction and oracle generation still remain. For example, many failures stem from expected values that do not match the current implementation's output and from incorrectly generated inputs.

For the \textit{Codec} project, \tech{} achieves higher line coverage than \pantavariant{} but slightly lower branch coverage. In particular, \tech{} improves line coverage by 3.99 pp (4.85\%), while reducing branch coverage by 3.99 pp (4.91\%) compared to \pantavariant{}. This result reflects the characteristics of the Codec project and the different mechanisms of the two approaches. Compared to other subjects, Codec contains a higher proportion of branch-intensive, rule-based logic, where execution paths are heavily determined by fine-grained input conditions. In such a setting, \tech{} benefits from call-chain-guided execution context construction, which helps resolve cross-class dependencies and improves reachability of executable code, leading to higher line coverage.

However, this mechanism tends to favor structurally guided execution paths, which may not sufficiently explore rare or highly conditional branches. In contrast, \pantavariant{} leverages uncovered-path feedback at the class level, which can better guide the exploration of less frequently exercised execution paths, thereby achieving slightly higher branch coverage.

This effect is also evident at the class level. For example, in the \textit{DoubleMetaphone} class, \pantavariant{} achieves 69.96\% line coverage and 57.53\% branch coverage, whereas \tech{} achieves 75.96\% line coverage and 57.08\% branch coverage. While \tech{} improves line coverage by 6.00 pp (8.58\%), it slightly decreases branch coverage by 0.45 pp (0.78\%).

These observations naturally raise the question of whether combining the complementary strengths of both approaches could further improve performance. Specifically, we investigate whether integrating uncovered-path information from \panta{} with the call-chain and dependency context from \tech{} yields additional gains. To answer this question, we conduct experiments on the Defects4J projects by augmenting \tech{}’s prompts with the uncovered-path information from \panta{}, resulting in the variant \techwithpath{} introduced in Section~\ref{sec:process}.

However, our results show that this combination does not consistently improve performance across projects. The detailed results (included in the replication package) show that \techwithpath{} achieves 71.30\% line coverage and 63.35\% branch coverage, while \tech{} still outperforms it by 10.71 and 9.88 pp (15.02\% and 15.60\%), respectively. One possible explanation is that including uncovered-path information substantially increases prompt length. In particular, in early iterations where branch coverage is still low, a large number of uncovered branches are included in the prompt. For branch-intensive focal methods, this can lead to prompt length increases of up to approximately 2×–3× compared to \tech{} alone. This enlarged prompt may dilute or overshadow critical call-chain and dependency signals, ultimately leading to suboptimal prompt effectiveness. We leave the exploration of more effective strategies for integrating these two complementary sources of information to future work.

\added{Although our work primarily focuses on improving code coverage, higher coverage is ultimately expected to enhance the defect-detection effectiveness of the generated tests. Therefore, to provide a more comprehensive evaluation of \tech{}, we additionally report the mutation scores achieved by \tech{}, \pantavariant{}, \panta{}, and EvoSuite in Table~\ref{tab:coverage}. On average, \tech{} achieves a mutation score of 67.80\%, outperforming \pantavariant{} (53.04\%), \panta{} (53.41\%), and EvoSuite (52.10\%) by 14.76, 14.39, and 15.70 pp (27.83\%, 26.94\%, and 30.13\%), respectively. These results show that \tech{} detects faults more effectively and highlight the benefits of incorporating call-chain and dependency contexts into LLM-based project-level test generation.} 

To assess the statistical significance of these improvements, we used non-parametric tests, Vargha and Delaney's $A_{12}$ statistic and the paired-sample Wilcoxon signed-rank test, as described in Section~\ref{sec:process}. Table~\ref{tab:stat_results} summarizes the results for both Defects4J and the unseen projects.
For RQ1 on Defects4J, \tech{} outperforms all baselines in terms of line coverage, branch coverage, and mutation score. Paired Wilcoxon signed-rank tests confirm that these improvements are statistically significant across all comparisons ($p<0.01$), well below the $\alpha=0.05$ threshold. The corresponding $A_{12}$ values range from 0.57 to 0.86, indicating large effect sizes in all comparisons except branch coverage against EvoSuite, for which the effect size is small (0.57).

\begin{table}[tp]
\centering
\scriptsize
\setlength{\tabcolsep}{3.5pt}
\renewcommand{\arraystretch}{1.1}
\caption{Statistical comparison of CAT against baselines on Defects4J and unseen projects.}
\label{tab:stat_results}
\begin{tabular}{llccccccccc}
\toprule
\multirow{2}{*}{\textbf{Dataset}} & \multirow{2}{*}{\textbf{Statistic}} & \multicolumn{3}{c}{\textbf{\tech{} vs \pantavariant{}}} & \multicolumn{3}{c}{\textbf{\tech{} vs \panta{}}} & \multicolumn{3}{c}{\textbf{\tech{} vs EvoSuite}} \\
\cmidrule(lr){3-5} \cmidrule(lr){6-8} \cmidrule(lr){9-11}
 & & \textbf{Line} & \textbf{Branch} & \textbf{Mut.} & \textbf{Line} & \textbf{Branch} & \textbf{Mut.} & \textbf{Line} & \textbf{Branch} & \textbf{Mut.} \\
\midrule
\multirow{2}{*}{Defects4J}
 & $p$-value & 0.00139 & 0.00056 & 0.00012 & 0.00003 & 0.00006 & 0.00024 & 0.00012 & 0.00216 & 0.00012 \\
 & $A_{12}$ effect size & 0.76 & 0.75 & 0.81 & 0.83 & 0.81 & 0.78 & 0.76 & 0.57 & 0.86 \\
\midrule
\multirow{2}{*}{Unseen Projects}
 & $p$-value & 0.00502 & 0.00667 & 0.00125 & 0.00896 & 0.00576 & 0.00025 & 0.00532 & 0.00497 & 0.00685 \\
 & $A_{12}$ effect size & 0.69 & 0.81 & 0.81 & 0.88 & 1.00 & 0.88 & 0.81 & 0.75 & 0.81 \\
\bottomrule
\end{tabular}%
\end{table}

\added{\textbf{Cost and generated tests.} Table~\ref{tab:cost} reports the LLM tokens consumed, the generation time, and the tests retained by the three LLM-based approaches on Defects4J. The time includes generation, repair, compilation, and execution; we report the one-time static analysis of \tech{} separately in Section~\ref{sec:process} (8.35 hours total, 2.4\% of its generation time). Summed over all projects, \tech{} consumes 154.9 million tokens and 353.8 hours (about 51K tokens and 7.1 minutes per focal method, or 2.72 hours per focal class), 2.09 and 2.99 times as much as \panta{} (74.1 million tokens and 118.5 hours), because it runs the generation-and-validation loop for each focal method rather than once per class. Across both datasets, \tech{} runs, per focal method, an average of 5 of the maximum \(N_{gen}=10\) generation iterations and 2 repair attempts (i.e., 7 LLM calls). LLM inference accounts for 27.16\% of the generation time, and compilation and test execution account for the rest. \pantavariant{}, which runs for the same time as \tech{}, consumes 228.8 million tokens and retains 18,405 tests (851 per project), 44.9\% more than \tech{} (12,702; 650 per project) and 65.9\% more than \panta{} (11,092; 549 per project); it retains more tests than \tech{} on 9 of the 14 projects. Yet \tech{} achieves 8.94, 9.61, and 14.76~pp (12.23\%, 15.11\%, and 27.83\%) higher line coverage, branch coverage, and mutation score (Table~\ref{tab:coverage}). Moreover, with these additional tests, \pantavariant{} improves line coverage over \panta{} by only 3.58~pp (5.15\%) and does not improve the mutation score (53.04\% vs.\ 53.41\%). The improvements of \tech{} are therefore unlikely to stem from the number of retained tests. Mocking frameworks are rarely used: in 0.0\% of the retained tests of \tech{} in every project except JacksonXml (2.9\%), and in at most 2.2\% of those of \pantavariant{} and \panta{} in any project, so the tests of all three approaches rarely rely on a mocking framework.} 

\added{In practice, when running \tech{}, the cost amounts to about 7.1 minutes and 51K tokens per focal method on our single-GPU server (Section~\ref{sec:implement}). Since test generation is typically performed offline rather than interactively, we consider this cost acceptable, in particular because it cannot be substituted by running \panta{} longer: with the same time budget, \pantavariant{} still achieves 8.94, 9.61, and 14.76~pp lower line coverage, branch coverage, and mutation score than \tech{}. The cost can also be reduced in practice: generation already stops early when coverage stagnates, most of the time is spent on compiling and executing tests rather than on LLM inference, and focal classes are processed independently and can thus be handled in parallel. Developers may also apply \tech{} selectively, e.g., to the focal methods that a cheaper tool such as EvoSuite leaves uncovered.}

\added{\textbf{Hallucinations in the generated tests.} On Defects4J, we observed two main types of hallucinations in the tests generated by \tech{}. \emph{API hallucinations} affected 13.8\% of all generated tests and involved references to nonexistent program elements or invalid uses of existing ones: accessing private or protected members (3.6\%), referencing nonexistent methods, fields, or constants (3.3\%), importing an existing class from the wrong package (2.9\%), invoking methods with incompatible argument types, mostly involving overloaded methods (1.9\%), misusing abstract classes or interfaces, e.g., by attempting to override a method that does not exist in the supertype (1.4\%), and referencing nonexistent classes or test helpers (0.7\%). \emph{Oracle hallucinations} affected 9.0\% of all generated tests and involved expected outcomes inconsistent with the program's observed behavior: incorrect asserted values or conditions (8.8\%) or expected exceptions that were not thrown (0.2\%). The remaining failures involved type mismatches, unhandled checked exceptions, runtime exceptions caused by invalid inputs or object states, and malformed output.}

\begin{rqbox}{RQ1 Summary}
On Defects4J, \tech{} achieves the highest average line coverage, branch coverage, and mutation score among all approaches. Compared with \pantavariant{}, \panta{}, and EvoSuite, it improves line coverage by 8.94, 12.52, and 9.67~pp (12.23\%, 18.02\%, and 13.37\%), branch coverage by 9.61, 12.97, and 4.17~pp (15.11\%, 21.52\%, and 6.04\%), and mutation score by 14.76, 14.39, and 15.70~pp (27.83\%, 26.94\%, and 30.13\%), respectively, showing that the call-chain and dependency contexts improve both test coverage and fault-detection capability. These gains come at about three times the generation time and twice the tokens of \panta{}. However, \pantavariant{}, which runs as long as \tech{} and uses more tokens while retaining more tests, still achieves lower coverage. At the same time, 13.8\% (API hallucinations) and 9.0\% (oracle hallucinations) of the tests generated on Defects4J initially fail because of hallucinations and must be repaired or discarded. Moreover, naively combining uncovered-path information with call-chain and dependency contexts does not lead to further improvements, suggesting that effectively integrating heterogeneous sources of information remains a significant challenge.
\end{rqbox}

\begin{table}[tbp]
\begin{addedblock}
\centering
\scriptsize
\setlength{\tabcolsep}{2.5pt}
\renewcommand{\arraystretch}{1.1}
\caption{Cost and characteristics of the generated tests across projects in Defects4J. Tok.: total LLM tokens consumed (millions); Time: end-to-end wall-clock time (hours); Mock: proportion of retained test methods that use a mocking framework (\%); \#Tests: number of retained tests. Average: geometric mean over projects for Tok., Time, and \#Tests; arithmetic mean for Mock.}
\label{tab:cost}
\begin{tabular}{l >{\columncolor{LightGray}}c >{\columncolor{LightGray}}c >{\columncolor{LightGray}}c >{\columncolor{LightGray}}c c c c c c c c c}
\toprule
\multirow{2}{*}{\textbf{Project}} & \multicolumn{4}{c}{\textbf{\tech{}}} & \multicolumn{4}{c}{\textbf{\pantavariant{}}} & \multicolumn{4}{c}{\textbf{\panta{}}} \\
\cmidrule(lr){2-5} \cmidrule(lr){6-9} \cmidrule(lr){10-13}
 & \textbf{Tok.} & \textbf{Time} & \textbf{Mock} & \textbf{\#Tests} & \textbf{Tok.} & \textbf{Time} & \textbf{Mock} & \textbf{\#Tests} & \textbf{Tok.} & \textbf{Time} & \textbf{Mock} & \textbf{\#Tests} \\
\midrule
JacksonXml & 6.0 & 19.5 & 2.9 & 167 & 8.8 & 19.5 & 0.0 & 262 & 1.6 & 4.2 & 0.0 & 123 \\
Csv & 3.7 & 4.4 & 0.0 & 364 & 5.7 & 4.4 & 1.6 & 272 & 3.9 & 2.9 & 0.0 & 227 \\
Collections & 10.8 & 26.1 & 0.0 & 812 & 22.2 & 26.1 & 0.0 & 1,486 & 5.5 & 6.9 & 0.0 & 817 \\
Gson & 3.8 & 5.4 & 0.0 & 195 & 6.0 & 5.4 & 0.0 & 603 & 4.0 & 3.5 & 0.0 & 517 \\
Cli & 1.6 & 1.4 & 0.0 & 194 & 1.2 & 1.4 & 0.0 & 131 & 0.7 & 0.9 & 0.0 & 94 \\
JacksonCore & 9.0 & 29.9 & 0.0 & 373 & 17.6 & 29.9 & 0.0 & 1,764 & 6.7 & 11.3 & 0.0 & 1,167 \\
JxPath & 6.4 & 12.4 & 0.0 & 748 & 7.4 & 12.4 & 0.0 & 893 & 7.1 & 11.7 & 0.0 & 889 \\
Jsoup & 8.8 & 19.7 & 0.0 & 1,258 & 10.3 & 19.7 & 0.0 & 801 & 3.8 & 7.3 & 0.0 & 512 \\
Codec & 3.9 & 5.1 & 0.0 & 741 & 3.4 & 5.1 & 0.0 & 465 & 2.5 & 3.1 & 0.0 & 207 \\
Compress & 3.6 & 10.2 & 0.0 & 425 & 4.3 & 10.2 & 0.0 & 374 & 4.1 & 9.9 & 0.0 & 388 \\
JacksonDatabind & 18.9 & 60.4 & 0.0 & 2,142 & 36.3 & 60.4 & 0.0 & 2,274 & 7.4 & 13.2 & 2.2 & 1,151 \\
Time & 22.7 & 36.6 & 0.0 & 1,620 & 23.3 & 36.6 & 0.0 & 2,089 & 5.3 & 9.0 & 0.0 & 1,107 \\
Lang & 31.3 & 56.1 & 0.0 & 1,833 & 46.7 & 56.1 & 0.0 & 4,059 & 9.9 & 12.6 & 0.0 & 2,086 \\
Math & 24.4 & 66.6 & 0.0 & 1,830 & 35.6 & 66.6 & 0.0 & 2,932 & 11.6 & 22.0 & 0.0 & 1,807 \\
\midrule
\textbf{Average} & 7.9 & 15.7 & 0.2 & 650 & 10.6 & 15.7 & 0.1 & 851 & 4.3 & 6.5 & 0.2 & 549 \\
\bottomrule
\end{tabular}
\end{addedblock}
\end{table}

\subsubsection{Impact of Call-Chain Extraction and Dependency Resolution.}\label{sec:rq2}

To investigate the contributions of call-chain awareness and dependency resolution in \tech{}, we compare \tech{} with three variants: \techvariant{}, which excludes both types of contextual information extracted in Section~\ref{sec:gen}; \techchain{}, which retains only call-chain context; and \techdep{}, which retains only dependency context. As shown in Table~\ref{tab:ablation}, \tech{} achieves 82.01\% line coverage, 73.23\% branch coverage, and a 67.80\% mutation score. In comparison, even the best-performing variant, \techdep{}, achieves only 66.30\% line coverage, 59.42\% branch coverage, and a 53.04\% mutation score. These results show that both types of contextual information contribute meaningfully and complementarily to test generation; retaining either alone substantially degrades performance. This complementarity stems from the different information they provide to the LLM. Call-chain context identifies feasible invocation paths to a focal method, whereas dependency context provides the information needed to construct and initialize the objects involved in those paths. For example, knowing an entry method in a call chain may be insufficient if the LLM lacks information about how to instantiate its declaring class or satisfy its dependencies. Conversely, dependency information alone may enable the construction of relevant objects but does not reveal how they can be composed into an invocation path that reaches the focal method. Combining the two therefore enables the LLM to both identify suitable invocation paths and construct the program states required to exercise them.

\begin{table}[tbp]
\centering
\scriptsize
\setlength{\tabcolsep}{3.5pt}
\renewcommand{\arraystretch}{1.1}
\caption{Ablation study on Defects4J: line coverage, branch coverage, and mutation score of \tech{} and its variants. \techvariant{}: without call-chain and dependency contexts; \techchain{}: call-chain context only; \techdep{}: dependency context only.}
\label{tab:ablation}
\begin{tabular}{l >{\columncolor{LightGray}}c >{\columncolor{LightGray}}c >{\columncolor{LightGray}}c c c c c c c c c c}
\toprule
\multirow{2}{*}{\textbf{Project}} & \multicolumn{3}{c}{\textbf{\tech{}}} & \multicolumn{3}{c}{\textbf{\techvariant{}}} & \multicolumn{3}{c}{\textbf{\techchain{}}} & \multicolumn{3}{c}{\textbf{\techdep{}}} \\
\cmidrule(lr){2-4} \cmidrule(lr){5-7} \cmidrule(lr){8-10} \cmidrule(lr){11-13}
& \textbf{Line} & \textbf{Branch} & \textbf{Mut.} & \textbf{Line} & \textbf{Branch} & \textbf{Mut.} & \textbf{Line} & \textbf{Branch} & \textbf{Mut.} & \textbf{Line} & \textbf{Branch} & \textbf{Mut.} \\
\midrule
JacksonXml & 64.16 & 58.34 & 61.85 & 56.41 & 56.07 & 54.18 & 45.65 & 51.23 & 53.26 & 59.03 & 56.86 & 56.94 \\
Csv & 82.60 & 70.83 & 79.60 & 59.25 & 49.70 & 47.32 & 63.40 & 53.40 & 53.90 & 62.87 & 52.90 & 62.01 \\
Collections & 92.90 & 90.05 & 86.28 & 61.50 & 55.66 & 53.12 & 62.80 & 59.52 & 59.30 & 66.36 & 61.51 & 62.79 \\
Gson & 87.73 & 75.07 & 68.87 & 69.30 & 57.24 & 54.43 & 71.32 & 60.93 & 56.02 & 71.97 & 60.47 & 57.85 \\
Cli & 87.98 & 78.47 & 81.81 & 66.74 & 49.44 & 62.16 & 65.29 & 49.27 & 61.49 & 69.56 & 53.58 & 65.49 \\
JacksonCore & 71.78 & 61.69 & 58.27 & 49.74 & 43.20 & 39.33 & 47.54 & 41.16 & 38.60 & 52.01 & 45.24 & 42.35 \\
JxPath & 74.13 & 66.26 & 53.08 & 57.58 & 63.04 & 43.23 & 55.11 & 62.56 & 39.57 & 59.01 & 64.17 & 45.35 \\
Jsoup & 88.45 & 71.90 & 63.33 & 75.54 & 64.06 & 51.00 & 76.54 & 64.11 & 56.54 & 76.59 & 64.08 & 55.63 \\
Codec & 86.28 & 77.20 & 59.64 & 80.82 & 77.46 & 46.50 & 81.28 & 76.65 & 46.20 & 81.58 & 76.95 & 47.57 \\
Compress & 64.22 & 59.52 & 51.86 & 51.42 & 42.31 & 33.52 & 49.50 & 39.76 & 32.27 & 50.59 & 41.62 & 32.46 \\
JacksonDatabind & 84.11 & 75.91 & 83.07 & 53.37 & 47.68 & 52.79 & 55.74 & 45.91 & 51.92 & 56.66 & 49.45 & 55.24 \\
Time & 92.28 & 84.36 & 74.47 & 81.68 & 75.60 & 57.06 & 81.12 & 75.58 & 58.55 & 80.59 & 74.84 & 57.70 \\
Lang & 86.48 & 78.25 & 71.75 & 65.85 & 60.64 & 53.12 & 70.16 & 65.67 & 53.87 & 70.33 & 65.88 & 53.87 \\
Math & 85.10 & 77.34 & 55.31 & 66.67 & 60.80 & 43.53 & 68.46 & 62.15 & 46.36 & 71.11 & 64.32 & 47.37 \\
\midrule
\textbf{Average} & 82.01 & 73.23 & 67.80 & 63.99 & 57.35 & 49.38 & 63.85 & 57.71 & 50.56 & 66.30 & 59.42 & 53.04 \\
\bottomrule
\end{tabular}
\end{table}

\added{\textbf{Precision of the extracted call chains.} Because \tech{} relies on CHA (Section~\ref{sec:cg-metadata}), some call chains passed to the LLM may be infeasible. We therefore examined all call paths that the static analysis passed into the prompts and identified those in which a virtual call resolves to an implementation incompatible with the receiver passed from the caller. Such spurious paths account for 2.41\% of all paths. A representative example occurs in Codec: \texttt{Soundex.\allowbreak difference} and \texttt{RefinedSoundex.\allowbreak difference} pass \texttt{this} to the package-private helper \texttt{SoundexUtils.\allowbreak difference}, which then calls \texttt{encode} on it. CHA resolves this call to every implementation of \texttt{StringEncoder}, including \texttt{DaitchMokotoffSoundex.\allowbreak encode}, and therefore presents \texttt{Soundex.\allowbreak difference} as an indirect entry point for the focal method \texttt{DaitchMokotoffSoundex.\allowbreak encode}, although no execution of \texttt{Soundex.\allowbreak difference} reaches it. Such paths add noise to the call-chain and dependency contexts, but they are rare, and a test built on a spurious entry point simply does not cover the focal method.}

\added{\textbf{Indirect invocation.} The call-chain context allows \tech{} to reach a focal method through other methods on its call chains, which is useful when the focal method is hard to invoke directly. Table~\ref{tab:indirect} quantifies the contribution of such indirect tests. We classify a retained test as indirect if it does not call the focal method in its body but executes it through a call chain, and we measure the gain attributable to indirect tests as the difference between the coverage and mutation score of the complete test suite and those of the same suite without the indirect tests. Across all projects, 25.9\% of retained \tech{} tests are indirect on average, and they contribute 18.5, 15.6, and 13.5~pp to line coverage, branch coverage, and mutation score, respectively. The share of indirect tests differs considerably between projects. In Csv, 34.0\% of the retained tests are indirect, and without them, line coverage, branch coverage, and mutation score would be lower by 17.1, 17.3, and 18.1~pp, respectively. In Collections, by contrast, only 1.6\% of the tests are indirect; they add 2.6~pp to line coverage but hardly affect branch coverage (0.5~pp) and mutation score (0.0~pp), likely because most of its focal methods belong to utility classes with public static methods, such as \texttt{IterableUtils}, \texttt{IteratorUtils}, and \texttt{MapUtils}, for which valid direct invocations can be easily constructed.}

\begin{table}[tbp]
\begin{addedblock}
\centering
\scriptsize
\setlength{\tabcolsep}{3.5pt}
\renewcommand{\arraystretch}{1.1}
\caption{Contribution of indirect tests in \tech{}. Indirect (\%): proportion of the retained tests that exercise the focal method through indirect calls along its call chains; Line, Branch, and Mut.: gain in line coverage, branch coverage, and mutation score (percentage points) attributable to these indirect tests.}
\label{tab:indirect}
\begin{tabular}{l r r r r}
\toprule
\textbf{Project} & \textbf{Indirect (\%)} & \textbf{Line (pp)} & \textbf{Branch (pp)} & \textbf{Mut. (pp)} \\
\midrule
JacksonXml & 15.4 & +5.7 & +5.2 & +6.4 \\
Csv & 34.0 & +17.1 & +17.3 & +18.1 \\
Collections & 1.6 & +2.6 & +0.5 & +0.0 \\
Gson & 37.2 & +29.3 & +27.4 & +24.8 \\
Cli & 25.0 & +8.9 & +6.0 & +5.6 \\
JacksonCore & 18.8 & +18.5 & +15.1 & +11.9 \\
JxPath & 34.4 & +25.6 & +9.2 & +7.6 \\
Jsoup & 35.7 & +26.7 & +24.0 & +13.0 \\
Codec & 31.1 & +37.5 & +28.8 & +25.6 \\
Compress & 34.1 & +23.3 & +23.0 & +22.8 \\
JacksonDatabind & 18.2 & +19.2 & +15.7 & +12.4 \\
Time & 21.8 & +22.3 & +28.2 & +21.8 \\
Lang & 35.6 & +28.7 & +22.5 & +14.2 \\
Math & 25.5 & +16.8 & +11.2 & +12.8 \\
\midrule
Adj & 33.2 & +21.7 & +5.2 & +15.9 \\
Astron & 24.3 & +7.5 & +14.3 & +9.3 \\
Binance & 23.1 & +15.5 & +12.7 & +10.0 \\
Joyagent & 17.9 & +6.7 & +13.6 & +10.8 \\
\midrule
\textbf{Average} & 25.9 & +18.5 & +15.6 & +13.5 \\
\bottomrule
\end{tabular}
\end{addedblock}
\end{table}

\begin{rqbox}{RQ2 Summary}
Call-chain and dependency contexts are complementary. On Defects4J, \tech{} improves line coverage, branch coverage, and mutation score over \techvariant{} by 18.02, 15.88, and 18.42~pp (28.16\%, 27.69\%, and 37.30\%), whereas using either context alone yields improvements of at most 2.31, 2.07, and 3.66~pp (\techdep{}). Across all 18 projects, 25.9\% of the retained \tech{} tests reach their focal methods indirectly through extracted call chains; on average, these indirect tests contribute 18.5, 15.6, and 13.5~pp to line coverage, branch coverage, and mutation score, respectively.
\end{rqbox}


\subsubsection{\tech{} Performance on Unseen Projects.} \label{sec:rq3}

As described in Section~\ref{sec:subjects}, we evaluate \tech{} on four projects that did not appear in the training data of the employed LLM and are more challenging, and compare its performance with \panta{} and \pantavariant{}. Table~\ref{tab:coverage_new} reports the line coverage, branch coverage, and mutation score results on these unseen projects. Across all projects, \tech{} consistently outperforms \panta{} and \pantavariant{}, improving the average line and branch coverage by 13.58 and 12.02 pp (25.09\% and 25.91\%), respectively, over \panta{}, and by 7.80 and 7.48 pp (13.02\% and 14.68\%) over \pantavariant{}. \added{\tech{} also achieves the highest mutation score on every unseen project, 50.75\% on average, compared with 30.87\% for \pantavariant{} and 26.11\% for \panta{}, i.e., 19.88 and 24.64~pp (64.40\% and 94.37\%) higher. Compared with EvoSuite, \tech{} improves the average line coverage, branch coverage, and mutation score by 23.19, 14.16, and 17.30~pp (52.10\%, 31.99\%, and 51.72\%) and achieves higher line coverage and mutation score on all four projects. On Binance, however, EvoSuite achieves much higher branch coverage than \tech{} (88.19\% vs.\ 67.70\%) but a lower mutation score (62.21\% vs.\ 65.97\%); as on Defects4J, higher coverage does not necessarily yield a higher mutation score.}

\begin{table}[tbp]
\centering
\scriptsize
\setlength{\tabcolsep}{3.5pt}
\renewcommand{\arraystretch}{1.1}
\caption{Line coverage, branch coverage, and mutation score across unseen projects. Bold values indicate the best performance per project.}
\label{tab:coverage_new}
\begin{tabular}{l >{\columncolor{LightGray}}c >{\columncolor{LightGray}}c >{\columncolor{LightGray}}c c c c c c c c c c}
\toprule
\multirow{2}{*}{\textbf{Project}} & \multicolumn{3}{c}{\textbf{\tech{}}} & \multicolumn{3}{c}{\textbf{\pantavariant{}}} & \multicolumn{3}{c}{\textbf{\panta{}}} & \multicolumn{3}{c}{\textbf{EvoSuite}} \\
\cmidrule(lr){2-4} \cmidrule(lr){5-7} \cmidrule(lr){8-10} \cmidrule(lr){11-13}
& \textbf{Line} & \textbf{Branch} & \textbf{Mut.} & \textbf{Line} & \textbf{Branch} & \textbf{Mut.} & \textbf{Line} & \textbf{Branch} & \textbf{Mut.} & \textbf{Line} & \textbf{Branch} & \textbf{Mut.} \\
\midrule
Adj & \textbf{66.06} & \textbf{56.59} & \textbf{\est{53.99}} & 60.67 & 49.90 & \est{31.00} & 52.78 & 43.63 & \est{25.46} & 43.17 & 34.17 & 29.63 \\
Astron & \textbf{59.33} & \textbf{52.64} & \textbf{\est{49.73}} & 51.24 & 43.70 & \est{12.11} & 50.07 & 43.34 & 12.52 & 21.55 & 20.04 & 14.24 \\
Binance & \textbf{85.27} & 67.70 & \textbf{\est{65.97}} & 74.82 & 58.14 & \est{59.77} & 65.47 & 51.29 & \est{51.85} & 74.91 & \textbf{88.19} & 62.21 \\
Joyagent & \textbf{60.15} & \textbf{56.73} & \textbf{\est{33.29}} & 52.87 & 52.02 & \est{20.61} & 48.17 & 47.33 & 14.61 & 38.42 & 34.64 & 27.72 \\
\midrule
\textbf{Average} & \textbf{67.70} & \textbf{58.42} & \textbf{\est{50.75}} & 59.90 & 50.94 & \est{30.87} & 54.12 & 46.40 & \est{26.11} & 44.51 & 44.26 & 33.45 \\
\bottomrule
\end{tabular}
\end{table}

Compared with Defects4J, the improvements over \panta{} on these projects are larger in relative terms (25.09\% and 25.91\% vs.\ 18.02\% and 21.52\%) and similar in absolute terms (13.58 and 12.02 pp vs.\ 12.52 and 12.97 pp), and the improvements over \pantavariant{} are comparable (7.80 and 7.48 pp vs.\ 8.94 and 9.61 pp; 13.02\% and 14.68\% vs.\ 12.23\% and 15.11\%), although these projects exhibit higher structural complexity. In contrast to Defects4J subjects, which are predominantly single-module repositories where classes are organized within a single module, the new projects are multi-module systems involving multiple source directories and more complex inter-module dependencies (as discussed in Section~\ref{sec:subjects}). As a result, successful test generation requires reasoning about more intricate dependency structures and execution contexts across modules, where the advantages of call-chain-guided dependency resolution in \tech{} become more pronounced.

Compared with RQ1, \tech{} achieves 82.01\% line coverage and 73.23\% branch coverage on Defects4J, while reaching 67.70\% and 58.42\% on unseen projects. This performance drop is likely attributable to the increased complexity of the unseen projects, which are more recent and exhibit a higher average maximum cyclomatic complexity (CYC$_{max}$) than the Defects4J subjects (33 vs.\ 27), and may also be influenced by potential data leakage effects in the Defects4J benchmark. Such projects often involve deeper call chains, richer third-party dependencies, and more intricate object initialization patterns, all of which make it more challenging to construct valid execution contexts.

Nevertheless, \tech{} consistently outperforms both LLM-based baselines across all unseen projects, confirming that the proposed call-chain- and dependency-aware context modeling strategy generalizes effectively beyond benchmark-specific settings. Importantly, because these projects postdate the employed LLM's cut-off, the observed gains cannot be explained by memorization and instead indicate genuine improvements in context construction and test generation.

The statistical analysis in Table~\ref{tab:stat_results} further reinforces this conclusion. Similar to RQ1, \tech{} demonstrates statistically significant improvements over both \pantavariant{} and \panta{} on unseen projects for both line and branch coverage ($p<0.01$). The corresponding $A_{12}$ values range from 0.69 to 1.00, indicating consistently medium-to-large effect sizes.

\added{\textbf{Cost and generated tests.} Table~\ref{tab:cost_new} reports the same measures for the unseen projects. \tech{} consumes 132.7 million tokens and 321.3 hours in total, 2.30 and 3.13 times as much as \panta{} (57.6 million tokens and 102.5 hours); Binance, which contains 2,335 of the 2,707 MUTs, accounts for 80.3\% of \tech{}'s time. 
\pantavariant{} consumes 235.5 million tokens in the same time as \tech{}. \tech{} retains 7,272 tests (673 per focal project), compared with 5,493 (474) for \pantavariant{} and 3,118 (393) for \panta{}. Unlike on Defects4J, \tech{} retains more tests than \pantavariant{} here (32.4\% more in total and more on three of the four projects), so a contribution of the test-suite size cannot be ruled out. However, with 76.2\% more tests than \panta{}, \pantavariant{} improves line coverage and mutation score over \panta{} by only 5.78~pp (10.68\%) and 4.76~pp (18.23\%), whereas \tech{} exceeds \pantavariant{} by 7.80~pp (13.02\%) and 19.88~pp (64.40\%), respectively. Moreover, on Joyagent, \tech{} retains fewer tests than both LLM-based baselines (117 vs.\ 124 and 163) yet achieves the highest coverage and mutation score (Table~\ref{tab:coverage_new}). As in Defects4J, mocking frameworks are rarely used, at most 2.3\% of the retained tests of \tech{} and 5.0\% of the baselines in any project.}

\begin{table}[tbp]
\begin{addedblock}
\centering
\scriptsize
\setlength{\tabcolsep}{2.5pt}
\renewcommand{\arraystretch}{1.1}
\caption{Cost and characteristics of the generated tests across unseen projects. Columns and Average as in Table~\ref{tab:cost}.}
\label{tab:cost_new}
\begin{tabular}{l >{\columncolor{LightGray}}c >{\columncolor{LightGray}}c >{\columncolor{LightGray}}c >{\columncolor{LightGray}}c c c c c c c c c}
\toprule
\multirow{2}{*}{\textbf{Project}} & \multicolumn{4}{c}{\textbf{\tech{}}} & \multicolumn{4}{c}{\textbf{\pantavariant{}}} & \multicolumn{4}{c}{\textbf{\panta{}}} \\
\cmidrule(lr){2-5} \cmidrule(lr){6-9} \cmidrule(lr){10-13}
& \textbf{Tok.} & \textbf{Time} & \textbf{Mock} & \textbf{\#Tests} & \textbf{Tok.} & \textbf{Time} & \textbf{Mock} & \textbf{\#Tests} & \textbf{Tok.} & \textbf{Time} & \textbf{Mock} & \textbf{\#Tests} \\
\midrule
Adj & 4.6 & 15.3 & 0.0 & 318 & 3.3 & 15.3 & 4.1 & 137 & 2.7 & 12.6 & 5.0 & 116 \\
Astron & 13.0 & 45.3 & 2.3 & 935 & 13.8 & 45.3 & 1.1 & 651 & 9.4 & 31.2 & 0.3 & 552 \\
Binance & 113.6 & 258.0 & 0.0 & 5,902 & 217.3 & 258.0 & 0.0 & 4,581 & 43.4 & 53.7 & 0.0 & 2,287 \\
Joyagent & 1.5 & 2.7 & 0.0 & 117 & 1.1 & 2.7 & 2.8 & 124 & 2.1 & 5.0 & 3.1 & 163 \\
\midrule
\textbf{Average} & 10.0 & 26.4 & 0.6 & 673 & 10.2 & 26.4 & 2.0 & 474 & 6.9 & 18.0 & 2.1 & 393 \\
\bottomrule
\end{tabular}
\end{addedblock}
\end{table}

At the same time, the lower absolute coverage on unseen projects suggests that project-level test generation for dependency-rich systems remains a challenging problem. Future work could explore more advanced techniques for modeling dynamic dependencies, improving context selection under long call chains, and integrating complementary signals beyond static call-chain and dependency information to further improve generalization.

\begin{rqbox}{RQ3 Summary}
On the four unseen projects (released after the employed LLM), \tech{} outperforms \panta{} and \pantavariant{} on every project and in every metric. On average, it improves line coverage by 13.58 and 7.80~pp (25.09\% and 13.02\%), branch coverage by 12.02 and 7.48~pp (25.91\% and 14.68\%), and mutation score by 24.64 and 19.88~pp (94.37\% and 64.40\%), respectively, at about three times the generation time of \panta{}. Compared with EvoSuite, \tech{} improves average line coverage, branch coverage, and mutation score by 23.19, 14.16, and 17.30~pp (52.10\%, 31.99\%, and 51.72\%), although EvoSuite achieves higher branch coverage on Binance. Because these projects are unlikely to be part of the LLM's training data, the gains are unlikely to be explained by memorization and can instead be attributed to \tech{}'s call-chain- and dependency-aware context modeling. Although absolute coverage decreases on these more complex projects, \tech{} maintains clear advantages over all baselines, demonstrating its robustness in realistic project-level test-generation scenarios.
\end{rqbox}

\subsection{Threats to Validity} \label{sec:threats}

\textbf{Internal validity.}
The performance of LLM-based approaches can be influenced by multiple factors, including the choice of LLM, prompt design, and configuration settings. To mitigate this threat, we adopt a state-of-the-art open-source LLM and use the same prompt design and configuration settings as \panta{} to ensure a fair comparison.

In addition, while \panta{} generates tests for all focal methods within a class in a single run, our approach \tech{} generates tests for each focal method individually. As discussed at the beginning of Section~\ref{sec:eval}, to ensure a fair comparison under this methodological difference, we also include a time-bounded variant of \panta{}, denoted \pantavariant{}, executed under the same time budget as \tech{}.

Another threat lies in the chosen parameter settings, such as \(N_{gen}\) and \(d_{max}\), which may affect both the effectiveness and efficiency of \tech{}. Although these values are determined through preliminary experiments, alternative configurations may lead to different performance trade-offs. However, because \pantavariant{} is evaluated under the same time budget, this threat is unlikely to invalidate the comparisons against the baseline, even if the absolute coverage numbers may vary. \added{Moreover, LLM outputs are non-deterministic, and because of the cost of the evaluation, each LLM-based approach was run once per focal class, so the results for some focal classes may differ between runs. However, our conclusions do not rest on individual classes: we analyze the results aggregated over all 242 focal classes of the 18 projects, which reduces the influence of run-to-run variation in individual classes on the overall comparison. In addition, \tech{} reduces its dependence on single LLM responses by generating tests over up to $N_{gen}$ iterations and retaining only tests that compile and pass; all LLM-based approaches use the same low temperature (0.2), and the results are aggregated over 242 focal classes.}

\added{Another threat concerns the precision of the static analysis. CHA over-approximates virtual dispatch and may thus introduce spurious call chains, which account for 2.41\% of the paths passed into the prompts (Section~\ref{sec:rq2}). Conversely, like other static call-graph analyses, it may miss calls made through reflection, dynamic proxies, or callbacks registered at run time; in addition, paths longer than $d_{max}$ are excluded to avoid path explosion, and entry points that are not publicly accessible are filtered out, although a test class in the package of the focal class could also call package-private and protected methods of this package (Section~\ref{sec:cg-metadata}). Since only 25.9\% of the retained tests of \tech{} reach their focal method through a call chain on average, this restriction affects only part of the generated tests; admitting such entry points and invoking package-private and protected focal methods directly may nevertheless improve coverage of some focal methods. Spurious paths add noise to the prompts, and missed paths reduce the indirect routes offered to the LLM, which may lower the coverage achieved for some focal methods.}

Nevertheless, different LLMs, prompting strategies, or parameter configurations may still yield different results. We therefore encourage future work to further investigate the generalizability of our findings across diverse LLMs, prompt designs, and configuration choices.

\textbf{External validity.} 
The main threat to external validity lies in the representativeness of the evaluated projects. To mitigate this concern, we conduct our evaluation on Java projects from Defects4J, a widely adopted benchmark in prior related research, as well as additional GitHub projects that have not been seen by the adopted LLM. Although this setup improves the diversity of the evaluation subjects, the findings may still not fully generalize to projects in other programming languages, domains, or industrial settings.

\added{Our evaluation uses Qwen3-Coder 30B, which was one of the strongest publicly available open-source models for coding tasks when this study was conducted. Different LLMs may respond differently to the same call-chain and dependency contexts. Therefore, the magnitude of the improvements we observe may vary across models, and our results should not be interpreted as model-independent improvements; rather, they support the potential of \tech{}. More capable models may also infer some project-level information from limited context, potentially reducing the relative improvements provided by \tech{}. Nevertheless, call-chain and dependency contexts explicitly provide project-specific information about invocation paths and dependencies that may not be available from the focal method alone. In particular, we did not evaluate closed-source models, as doing so would incur substantial additional costs. Prior studies do not report a consistent advantage of closed over open models for unit test generation: Yang et al.~\citep{yang2024evaluation} found that GPT-4 achieved higher line coverage than open models with up to 34B parameters, whereas one of the open models achieved higher branch coverage, and in the evaluation of \panta{}~\citep{gu2025llm}, an open model achieved higher coverage than one closed model and lower coverage than another. Evaluating \tech{} across different LLM families and capabilities remains an important direction for future work.}

\textbf{Construct validity.} 
The main threat to construct validity is potential overlap between the LLM's training data and the evaluated projects, which may inflate performance. To mitigate this issue, we further evaluate our approach on projects that are unlikely to have appeared in the LLM’s training data. As described in Section~\ref{sec:subjects}, we select four Java projects from GitHub that were released after the LLM’s release date and have more than 500 stars. The consistent results on these unseen projects further confirm the effectiveness of our approach, suggesting that the observed performance gains are not solely attributable to memorization.

\section{Related Work} \label{sec:related}
\subsection{Recent LLM-Based Test Generation with Execution Feedback}
Recent advances in LLM-based test generation have demonstrated promising results in producing high-quality unit tests. Representative approaches include \panta{}~\citep{gu2025llm}, SymPrompt~\citep{ryan2024code}, and PALM~\citep{wu2025generating}, the former being the current state of the art. These methods primarily rely on execution-path information to guide LLM-based test generation. However, as discussed in Section~\ref{sec:motivate}, path-based guidance alone is often insufficient for complex software systems with rich inter-class dependencies and intricate object initialization requirements.

In contrast, our approach \tech{} explicitly incorporates call-chain and dependency information into prompt design through static analysis, enabling the LLM to construct valid execution contexts for focal methods in complex projects. We evaluate \tech{} on both the Defects4J benchmark, used for evaluating existing approaches, and four additional GitHub projects released after the cutoff date of the employed LLM. Experimental results show that \tech{} consistently outperforms the state-of-the-art path-guided approach (\panta{}). We also investigate combining path execution information with our approach, but find that it does not yield improvements across projects (Section~\ref{sec:rq1}), suggesting that naive integration of complementary signals is non-trivial.

\subsection{Program-Analysis-Augmented Test Generation} \label{sec:related-pa}
\added{Traditional automated test generation explores the program under test with program analysis rather than LLM reasoning. Random testing, as in Randoop~\citep{pacheco2007randoop}, extends sequences of method calls using execution feedback; search-based testing, as in EvoSuite~\citep{fraser2011evosuite}, evolves test suites towards coverage goals with metaheuristic search; and symbolic execution, as in KLEE~\citep{cadar2008klee} and Pex~\citep{tillmann2008pex}, derives test inputs by solving path constraints~\citep{cadar2013symbolic}. These tools construct the objects and arguments needed to reach the code under test themselves, through random or evolved call sequences or through constraint solving, and derive their oracles from observed executions or from generic properties such as crashes. Among them, Randoop and EvoSuite target Java unit tests, the setting of this paper, and EvoSuite serves as the non-LLM baseline of our evaluation (Section~\ref{sec:eval}). LLM-based approaches, including \tech{}, instead rely on the context in the prompt to construct the required objects and to invoke the focal method, which is the problem that the call-chain and dependency contexts of \tech{} address.}

A line of recent LLM-based approaches also leverages static analysis to provide additional context for test generation. For example, CityWalk~\citep{zhang2025citywalk} incorporates project-level dependency awareness for C++ test generation. Roychowdhury et al.~\citep{roy2024static} extract method signatures, declaring classes, and field type information to enrich prompts. KAT~\citep{le2024kat} constructs operation-level dependency graphs for REST API testing, while RUG~\citep{cheng2025rug} uses type dependency graphs for Rust test generation. HITS~\citep{wang2024hits} and TELPA~\citep{yang2024enhancing} apply program slicing and contextual analysis to improve coverage of hard-to-reach branches. 
PAINT~\citep{nan2025test} explicitly targets branch coverage by performing control-flow graph analysis to extract condition chains that guide the LLM toward executing specific branches. In addition, it trains a machine learning model to predict whether a method call can be replaced with a mock, thereby supporting test construction when direct invocation is not feasible.
ASTER~\citep{pan2025aster} adopts a lightweight program analysis strategy to support multi-language test generation. For Java, it focuses on constructing executable test scaffolds by identifying class fields whose types correspond to mockable APIs, and then generating prompts that include constructors, accessor methods (i.e., getters and setters), and method stubs. This design primarily enables object instantiation and external dependency handling via mocking to ensure test executability, rather than explicitly guiding branch exploration.
Importantly, these two approaches operate at different abstraction levels and serve different purposes than \tech{}: PAINT focuses on branch-level execution guidance, while ASTER focuses on test scaffolding for ensuring executability.
In comparison, \tech{} performs call-chain extraction and dependency resolution at a more holistic level by integrating call chains, object initialization information, and inter-module dependencies, rather than focusing solely on branch-specific guidance or test scaffolding to ensure executability.
Notably, ASTER and PAINT do not publicly release their implementations, which limits reproducibility and direct comparison in our experiments.

Besides generation-time analysis, another line of work explores using static analysis as a post-processing step. For instance, LLMLOOP~\citep{ravi2025llmloop} and Dolcetti et al.~\citep{dolcetti2026helping} employ static analysis to filter invalid tests or guide iterative repair after generation.

A recent survey on LLM-based test generation~\citep{zhang2026enhancing} highlights that major challenges remain in handling complex software systems, mitigating hallucinations, and improving test validity. It further identifies program analysis, particularly static analysis, as a promising direction for enhancing LLM-based test generation. In this paper, we contribute to that research direction. 

In contrast to existing work, \tech{} focuses on leveraging call-chain and dependency contexts and integrating them into LLM prompts. To the best of our knowledge, this is the first work to explicitly investigate the role of such context information in project-level LLM-based test generation and demonstrate its effectiveness on both a widely used benchmark and complex real-world software systems.
\section{Conclusion} \label{sec:conclusion}
In this paper, we presented \tech{}, an LLM-based test generation approach built on a dedicated static analysis framework that systematically captures execution contexts. By explicitly modeling call-chain and dependency contexts---including caller--callee relationships, object constructors, and third-party dependencies---\tech{} helps the LLM construct executable and semantically valid contexts for project-level test generation. Our evaluation shows that \tech{} on average outperforms state-of-the-art approaches on both Defects4J (a widely used benchmark in prior work) and real-world GitHub projects released after the employed LLM's cut-off date.

We also find that naively combining execution-path information with other signals does not necessarily improve performance, suggesting that effectively integrating complementary sources of information remains challenging. Our ablation study further shows that both call-chain and dependency contexts are essential components of \tech{}.

Overall, our findings highlight the importance of modeling call-chain-aware execution contexts in LLM-based test generation. We suggest that future work explore more principled methods for integrating heterogeneous signals (e.g., execution paths and input--output relationships) to better support automated test generation. \added{Future work could also improve the precision and completeness of the extracted contexts. For example, propagating receiver types along call chains could prune incompatible dispatch targets, and comparing extracted paths with call edges observed during test execution could quantify how many relevant paths are missed. Since generated tests reside in the focal class's package, we could admit package-private and protected methods in that package as entry points, and invoke package-private and protected focal methods directly; this would require adapting the visibility filter, the initialization context, and the generation prompt.}

\added{In addition, a finer-grained analysis of test fixing across iterations, together with more effective fixing strategies, could improve the recovery of failed generated tests. \tech{} could also start from an existing test suite (e.g., one generated by EvoSuite) and generate tests only for code left uncovered, potentially reducing the time and cost of test generation. Like the compared approaches, \tech{} generates regression tests whose assertions reflect current program behavior; combining \tech{} with oracle generation techniques~\citep{dinella2022toga,jahangirova2016test} could yield tests that also check intended behavior. Finally, evaluating \tech{} with additional open and closed LLMs, and repeating experiments to measure variance across runs, would clarify the extent to which its improvements depend on the model and on the randomness of LLM sampling.}
\section*{Acknowledgements}

We sincerely thank the authors of \panta{} for making their implementation publicly available and for their assistance in clarifying experimental details during replication.

This work has emanated from research jointly funded by Taighde Éireann -- Research Ireland under Grant No.~13/RC/2094\_2 and by Huawei Technologies Co., Ltd. Lionel Briand is also supported by the Natural Sciences and Engineering Research Council of Canada.

For the purpose of Open Access, the authors have applied a CC BY public copyright licence to any Author Accepted Manuscript version arising from this submission.

\bibliographystyle{ACM-Reference-Format}
\bibliography{sample-base}


\end{document}